\documentstyle[aps,amstex,amssymb,eqsecnum,epsf]{revtex}

\begin{document}

\title{Boundary conditions in linearized harmonic gravity}

\author{
	B\'{e}la Szil\'{a}gyi${}^{1}$ ,
        Bernd Schmidt${}^{2}$,
	Jeffrey Winicour${}^{1,2}$
       }
\address{
${}^{1}$ Department of Physics and Astronomy \\
         University of Pittsburgh, Pittsburgh, PA 15260, USA \\
${}^{2}$ Max-Planck-Institut f\" ur
         Gravitationsphysik, Albert-Einstein-Institut, \\
	 14476 Golm, Germany
	 }
\date{\today}
\maketitle

\begin{abstract}

We investigate the initial-boundary value problem for linearized gravitational
theory in harmonic coordinates. Rigorous techniques for hyperbolic systems are
applied to establish well-posedness for various reductions of the system into a
set of six wave equations. The results are used to formulate computational
algorithms for Cauchy evolution in a 3-dimensional bounded domain. Numerical
codes based upon these algorithms are shown to satisfy tests of robust
stability for random constraint violating initial data and random boundary
data; and shown to give excellent performance for the evolution of typical
physical data. The results are obtained for plane boundaries as well as
piecewise cubic spherical boundaries cut out of a Cartesian grid.

\end{abstract}


\section{Introduction}

Grid boundaries pose major difficulties in current computational efforts to
simulate 3-dimensional black holes by conventional Cauchy evolution schemes.
The  initial-boundary value problem for Einstein's equations consists of the
evolution of initial Cauchy data on a spacelike hypersurface and boundary data
on a timelike hypersurface. This problem has only recently received
mathematical attention. Friedrich and Nagy \cite{Friedrich98} have given a
full solution for a hyperbolic formulation of the Einstein equations based
upon a frame decomposition in which the connection and curvature enter as
evolution variables. Because this formulation was chosen to handle
mathematical issues rather than for ease of numerical implementation, it is
not clear how the results translate into practical input for formulations on
which most computational algorithms are based.  The proper implementation of
boundary conditions depends on the particular reduction of Einstein's
equations into an evolution system and the choice of gauge conditions. The
purpose of this paper is to elucidate, in a simple context, such elementary
issues as: (i) which variables can be freely specified on the boundary; (ii)
how should the remaining variables be updated in a computational scheme; (iii)
how can the analytic results be implemented as a computational algorithm. For
this purpose, we consider the evolution of the linearized Einstein's equations
in harmonic coordinates \cite{dedonder,lanczos} and demonstrate how a robustly
stable and highly accurate computational evolution can be based upon a proper
mathematical formulation of the initial-boundary value problem.

Harmonic coordinates were used to obtain the first hyperbolic formulation of
Einstein's equations \cite{bruhat}. See Ref. \cite{Friedrich00} for a full
account of hyperbolic formulations of general relativity. While harmonic
coordinates have also been widely applied in carrying out analytic perturbation
expansions, they have had little application in numerical relativity,
presumably because of the restriction in gauge freedom \cite{bona}. However,
there has been no ultimate verdict on the suitability of harmonic coordinates
for computation.  In particular, their generalization to include gauge source
functions \cite{friedsourc} appears to offer flexibility comparable to the
explicit choice of lapse and shift in conventional numerical approaches to
general relativity. There is no question that harmonic coordinates offer
greater computational efficiency than any other hyperbolic formulation. See
Ref. \cite{Garf} for a recent application to the study of singularities in a
space-time without boundary. Here we use a reduced version of the harmonic
formulation of the field equations. This allows us retain a symmetric
hyperbolic system, and apply standard boundary algorithms, in a way that is
consistent with the propagation of the constraints. We show, on an analytic
level, that this leads to a well posed initial-boundary problem for linearized
gravitational theory.

Our computational results are formulated in terms of a Cartesian grid based
upon background Minkowskian coordinates. Robustly stable evolution algorithms
are obtained for plane boundaries aligned with the Cartesian coordinates, which
is the standard setup for three dimensional evolution codes. Similar
computational results \cite{cboun}, based upon long evolutions with random
initial and boundary data, were previously found for the linearized version of
the Arnowitt-Deser-Misner formulation (ADM) \cite{adm} of the field equations,
where the lack of a hyperbolic formulation required a less systematic approach
which had no obvious generalization to other boundary shapes. In this paper, we
also attain robust stability for spherical boundaries which are cut out of the
Cartesian grid in an irregular piecewise cubic fashion. This success gives
optimism that the methods can applied to such problems as black hole excision
and Cauchy-characteristic matching, where spherical boundaries enter in a
natural way.

Conventions: We use Greek letters for space-time indices and Latin letters for
spatial indices, e.g. $x^\alpha=(t,x^i)$ for standard Minkowski coordinates.
Linear perturbations of a curved space metric $g_{\alpha\beta}$ about the
Minkowski metric $\eta_{\alpha\beta}$ are described by  $\delta
g_{\alpha\beta}= h_{\alpha\beta}$ with similar notation for the corresponding
curvature quantities, e.g. the linearized Riemann tensor $\delta
R_{\alpha\beta\gamma\delta}$ and linearized Einstein tensor $\delta
G_{\alpha\beta}$. Indices are raised and lowered with the Minkowski metric,
with the result that $\delta g^{\alpha\beta}=-h^{\alpha\beta}$. Boundaries in
the background geometry are described in the form $z=const$ with the spatial
coordinates decomposed in the form $x^i=(x^A,z)$, where the $x^A=(x,y)$
directions span the space tangent to the boundary. 

\section{Harmonic evolution without boundaries}

The linearized Einstein tensor has the form
\begin{equation}
   \delta G^{\alpha\beta}=-\frac{1}{2}\Box \gamma^{\alpha\beta} 
        - \partial^{(\alpha}\Gamma^{\beta)}
	+\frac{1}{2}\eta^{\alpha\beta}\partial_\mu \Gamma^\mu ,
\label{eq:einstein}
\end{equation}
where $\Box =\partial_\mu \partial^\mu$
\begin{equation}
     \gamma^{\alpha\beta}=h^{\alpha\beta}-\frac{1}{2}\eta^{\alpha\beta} h
\end{equation}
$h=h^\alpha_\alpha =-\gamma^\alpha_\alpha =-\gamma$ and
\begin{equation}
     \Gamma^{\alpha}=-\partial_\beta \gamma^{\alpha\beta}.
\end{equation}
We analyze these equations in standard background Minkowskian coordinates
$x^\alpha$. To linearized accuracy, 
\begin{equation}
        \Gamma^{\alpha}\approx g^{\mu\nu}\nabla_\mu \nabla_\nu x^\alpha
\end{equation}
in terms of the curved space connection $\nabla_\mu$ associated with
$g_{\alpha\beta}$ and the condition $\Gamma^\alpha =0$ defines
a linearized harmonic gauge.

The diffeomorphisms of the curved spacetime induce equivalent metric
perturbations according to the harmonic subclass of gauge transformations
\begin{equation}
       \hat h^{\alpha\beta}= h^{\alpha\beta}
	  +2\partial^{(\alpha}\xi^{\beta)},
\label{eq:gauge}
\end{equation}
where the vector field $\xi^\alpha$ satisfies $\Box \xi^\alpha =0$.
The linearized curvature tensor
\begin{equation}
    2\delta R_{\mu\alpha\nu\beta}=\partial_\mu \partial_\beta h_{\alpha\nu}
                     +\partial_\alpha \partial_\nu h_{\mu\beta}
		     -\partial_\alpha \partial_\beta h_{\mu\nu}
		     -\partial_\mu \partial_\nu h_{\alpha\beta},
\label{eq:curv}
\end{equation}
as well as the linearized Einstein equations, are gauge invariant. 

We introduce a Cauchy foliation $t$ in the perturbed spacetime
such that it reduces to an inertial time slicing
in the background Minkowski spacetime. The unit normal to the
Cauchy hypersurfaces is given, to linearized accuracy, by
\begin{equation}
       n_\alpha\approx -(1+\frac{1}{2}h^{tt})\partial_\alpha t .
\end{equation}
The choice of an evolution direction $t^\mu=\alpha n^\mu +\beta^\mu$, with
unit lapse and shift in the background Minkowski spacetime, defines a
perturbative lapse $\delta \alpha =-h^{tt}/2$ and perturbative shift $\delta
\beta^i = -h^{ti}$.

\subsection{Standard harmonic evolution}

Harmonic evolution consists of solving the Einstein's equations 
$G_{\alpha\beta}=0$, subject to the harmonic conditions $\Gamma^\alpha =0$.
This formulation led to the first existence and uniqueness theorems for
solutions to the nonlinear Einstein equations by considering them as a set of
10 nonlinear wave equations \cite{choqbru}.

In the linearized case, Einstein's equations in harmonic coordinates reduce to
ten flat space wave equations so that their mathematical analysis is simple.
The Cauchy data $\gamma^{\alpha\beta}(0,x^i)$ and
$\partial_t\gamma^{\alpha\beta}(0,x^i)$ at $t=0$ determine ten unique solutions
$\gamma^{\alpha\beta}(t,x^i)$ of the wave equation $\Box
\gamma^{\alpha\beta}=0$, in the appropriate domain of dependence. These
solutions satisfy $\Box \Gamma^\alpha=0$ so that they satisfy Einstein's
equations provided $\Gamma^\alpha(0,x^i)=0$ and $\partial_t
\Gamma^\alpha(0,x^i)=0$, which can be arranged by choosing initial Cauchy data
satisfying constraints.  For a detailed discussion, see Ref. \cite{wald1984}. 

Although this standard harmonic evolution scheme led to the first existence and
uniqueness theorem for Einstein's equations, it is not straightforward to apply
to the initial-boundary value problem. The ten wave equations for
$\gamma^{\alpha\beta}$ require ten individual pieces of boundary data in order
to determine a unique solution. Given initial data such that  $\Gamma^\alpha=0$
and $\partial_t \Gamma^\alpha=0$ at $t=0$, as described above, the resulting
solution satisfies the linearized Einstein equations only in the domain of
dependence of the Cauchy data. In order that the solution of Einstein's
equations extend to the boundary, it is necessary that $\Gamma^\alpha=0$ at the
boundary. Unfortunately, there is no version of boundary data for
$\gamma^{\alpha\beta}$, e.g. Dirichlet, Neumann or Sommerfeld data for the ten
individual components, from which $\Gamma^\alpha$ can be calculated at the
boundary. Here we consider reduced versions of the harmonic evolution scheme in
which only six wave equations are solved and this problem does not arise. These
reduced harmonic formulations are presented below.  

\subsection{Reduced harmonic Einstein evolution}

A linearized evolution scheme for the harmonic Einstein system, can be based
upon the six wave equations
\begin{equation}
    \Box \gamma^{ij} =0
    \label{eq:gammawave} 
\end{equation}
along with the four harmonic
conditions
\begin{equation}
    \partial_\mu \gamma^{\mu\nu} = 0.
    \label{eq:auxeqs}
\end{equation}
Because of the harmonic conditions, this system satisfies the spatial
components $\delta G^{ij}=0$ of the linearized Einstein's equations.

As  a result of the linearized Bianchi identities $\partial_\alpha \delta
G^{\alpha\beta}=0$,  or
\begin{eqnarray}
        \partial_t \delta G^{it}+\partial_j \delta G^{ij} &=& 0 \\
	\partial_t \delta G^{tt}+\partial_j \delta G^{tj} &=& 0,
\end{eqnarray}
the linearized Hamiltonian constraint ${\cal C}:= \delta G^{tt}=0$ and
linearized momentum constraints ${\cal C}^i:=\delta G^{ti}=0$ are also
satisfied, throughout the domain of dependence of the Cauchy data,
provided that they are satisfied at the initial time
$t=0$. This constrains the initial values of
$\gamma^{t\alpha}$ according to
\begin{equation}
    \nabla^2 \gamma^{ti}+\partial_t\partial_j\gamma^{ij} =0 
\label{eq:auxil1}
\end{equation}
and
\begin{equation}
    \nabla^2 \gamma^{tt}-\partial_i\partial_j\gamma^{ij} =0,
\label{eq:auxil2}
\end{equation}
where $\nabla^2 =\delta^{ij} \partial_i \partial_j$.
Then, if these constraints are initially satisfied, the reduced harmonic
Einstein system determines a solution of the linearized Einstein's equations.

The well-posedness of the system follows directly from the well-posedness of
the wave equations for $\gamma^{ij}$. The auxiliary variables $\gamma^{\alpha
t}$ satisfy the ordinary differential equations
\begin{eqnarray}
        \partial_t \gamma^{it}+\partial_j \gamma^{ij} &=& 0 
	\label{eq:einstaux1}\\
	\partial_t \gamma^{tt}+\partial_j \gamma^{tj} &=& 0,
	\label{eq:einstaux2}
\end{eqnarray}
where $\gamma^{ij}$ enters only in the role of source terms. These
differential equations do not affect the well-posedness of the system and have
unique integrals determined by the initial values of $\gamma^{\alpha t}$. 

\subsection{Reduced harmonic Ricci evolution}

The reduced harmonic Ricci system consists of the six wave equations
\begin{equation}
      \Box h^{ij}=0
      \label{eq:hwave}
\end{equation}
along with the four harmonic conditions (\ref{eq:auxeqs}),
which can be re-expressed in the form
\begin{eqnarray}
\label{eq:gauge_t}
     &&  \partial_t \phi + \partial_j h^{jt}
      + \frac{1}{2} \, \partial_t h^j_j = 0 ,\\
     \label{eq:gauge_i}
      &&  \partial^i \phi + \partial_t h^{i t} 
    - \frac{1}{2} \,  \partial^i h^j_j 
   + \partial_j h^{i j} = 0 ,
\end{eqnarray}
where we have set $\phi = h^{tt}/2$. Together these equations imply that the
spatial components of the perturbed Ricci tensor vanish, $\delta R^{ij}=0$. In
addition, the Bianchi identities imply that the remaining components satisfy
\begin{eqnarray}
        \partial_t {\cal C}^i+\partial^i {\cal C}+\partial_j \delta R^{ij} &=& 0
	\label{eq:momprop} \\
	\partial_t {\cal C}+\partial_j {\cal C}^j &=& 0,
	\label{eq:hamprop}
\end{eqnarray}
where, in terms of the Ricci tensor, ${\cal C}=\frac{1}{2}(\delta R^{tt}+\delta
R^i_i)$ and ${\cal C}^i=\delta R^{ti}$. Together with the evolution equations
$\delta R^{ij}=0$, the Bianchi identities imply that the Hamiltonian constraint
satisfies the wave equation $\Box {\cal C}=0$.

If the Hamiltonian and momentum constraints are satisfied at the initial
time, then $\partial_t {\cal C}$ also vanishes at the initial time so that
the uniqueness of the solution of the wave equation ensures the propagation
of the Hamiltonian constraint. In turn, Eq.~(\ref{eq:momprop}) then ensures
that the momentum constraint is propagated. Thus the reduced harmonic
Ricci system of six wave equations  and four harmonic equations leads to a
solution of the linearized Einstein's equations for initial Cauchy data
satisfying the constraints.

The harmonic Ricci system takes symmetric hyperbolic form when the wave
equations are recast in first differential order form. Thus the system is
well posed.  The formulation of the harmonic Ricci system as a symmetric
hyperbolic system and the description of its characteristics is given in
Appendix \ref{app:ricci}. Although well-posedness of the analytic problem does
not the guarantee the stability of a numerical implementation it can simplify
its attainment. 

\subsection{Other reduced harmonic systems}

The harmonic Einstein and Ricci systems are special cases of a one parameter
class of reduced harmonic systems for the variable $$\kappa^{\alpha\beta}=
\gamma^{\alpha\beta}-\frac{\lambda}{2}\eta^{\alpha\beta}\gamma$$ which
satisfying the wave equations $E^{ij}=-\frac{1}{2} \Box \kappa^{ij}= 0$ and  the
harmonic conditions $\Gamma^\alpha=0$, where $E^{\alpha\beta}=\delta
G^{\alpha\beta} -\frac{\lambda}{2}\eta^{\alpha\beta}\delta G$. This system is
symmetric hyperbolic when the auxiliary system $\Gamma^\alpha=0$ is symmetric
hyperbolic. This can be analyzed by setting $\kappa^{ij}= 0$ in the auxiliary
system, which then takes the form 
\begin{eqnarray}
     0 &=& \frac{(2-3\lambda)}{2(1-2\lambda)}\partial_t\kappa^{tt}
         +\partial_j\kappa^{tj} \nonumber \\
     0 &=& \partial_t\kappa^{ti}
          -\frac{\lambda}{2(1-2\lambda)}\partial^i \kappa^{tt},	 
\end{eqnarray}
and implies
\begin{equation}
   (\partial_t^2 -\frac {\lambda}{(3\lambda-2)}\partial_i\partial^i)
          \kappa^{tt}  =0.
	  \label{eq:subwave}
\end{equation}
The auxiliary system is symmetric hyperbolic when Eq.~(\ref{eq:subwave})
is a wave equation whose wave speed
\begin{equation}
         v=\sqrt{\frac{\lambda}{(3\lambda-2)}}
\end{equation}
is positive. This is satisfied for $\lambda <0$ and $\lambda >2/3$. In the
range $2/3<\lambda < 1$, the wave speed is faster than the speed of light. Only
for the case $\lambda=1$, the harmonic Ricci system, is the wave speed of the
auxiliary system equal to the speed of light.

The auxiliary system for the reduced harmonic Einstein case has a well posed
initial-boundary value problem but represents a borderline case. This could
adversely affect the development of a stable code based upon a nonlinear
version of the reduced harmonic Einstein system.

\section{The initial-boundary value problem}

Consider the initial-boundary problem consisting of evolving Cauchy data
prescribed in a set ${\cal S}$ at time $t=0$ lying in the half-space $z>0$ with
data prescribed on a set ${\cal T}$ on the boundary $z=0$. Harmonic evolution
takes its simplest form when Einstein equations are expressed as a second
differential order system. However, in order to apply standard methods it is
necessary to recast the problem as a first order symmetric hyperbolic system.
Then the theory determines conditions on the boundary data for a well posed
problem in the future domain of dependence ${\cal D}^+$. Here ${\cal D}^+$ is
the maximal set of points whose past directed inextendible characteristic
curves all intersect the union of ${\cal S}$ and ${\cal T}$ before leaving
${\cal D}^+$.

Appendix \ref{app:ricci} describes the symmetric hyperbolic description of
boundary conditions for the 3-dimensional wave equation. The basic ideas and
their application to the initial-boundary value problem are simplest to explain
in terms of the 1-dimensional wave equation, as follows. See  Appendix
\ref{app:ricci} for the analogous treatment of the 3-dimensional case. Our
presentation is based upon the formulation of maximally dissipative boundary
conditions \cite{rauch}, the approach used by Friedrich and Nagy
\cite{Friedrich98} in the nonlinear case. An alternative description of
boundary conditions for symmetric hyperbolic systems is given in Ref.
\cite{kreiss} and for linearized gravity in Ref. \cite{stewart}.

\subsection{The 1-dimensional wave equation}
The one-dimensional wave equation $h_{,tt} = h_{,zz}$
can be recast as the first order system of evolution equations
\begin{eqnarray}
\label{eq:swe.first}
\dot h &=& k \\
\dot k &=& f_{,z} \\
\dot f &=& k_{,z},
\label{eq:swe.last}
\end{eqnarray}
where we have introduced the auxiliary variables $k$ and $f$.
Given initial data $(h,k,f)$ at time $t=0$ subject to the constraint,
$f=h_{,z}$, these equations determine a solution $h$ of the wave equation.
Note that the constraint is propagated by the evolution equations.
  
In coordinates $x^\mu=(x^0,x^1)=(t,z)$
the system (\ref{eq:swe.first})-(\ref{eq:swe.last})
has the symmetric hyperbolic form 
\begin{equation}
A^\mu \partial_\mu u = B u + F ,
\label{shscalar}
\end{equation}
where the solution $u$ consists of the column matrix
\begin{equation}
u = \left( \; \begin{array}{l} h \\ k \\ f \end{array} \right) 
\end{equation}
and
\begin{equation}
A^0 = \left( \; 
\begin{array}{ccc} 
1 & 0 & 0 \\
0 & 1 & 0 \\
0 & 0 & 1 
\end{array} \right) , \quad 
A^1 = \left( \; 
\begin{array}{ccc} 
0 & 0 & 0 \\
0 & 0 & -1 \\
0 & -1 & 0 
\end{array} \right) , \quad
B = \left( \; 
\begin{array}{ccc} 
0 & 1 & 0 \\
0 & 0 & 0 \\
0 & 0 & 0 
\end{array} \right) .
\end{equation}
In our case the source term $F=0$  but otherwise it plays no essential role in
the analysis of the system.

The contraction of Eq.~(\ref{shscalar}) with the transpose ${}^T u$ give
the flux equation 
\begin{equation}
  {}^T u A^\mu \partial_\mu u ={}^T u({}^T B +B)u.
\end{equation}
This can be used to provide an estimate on the norm $\int {}^T u A^0 u dz$ for
establishing a well posed problem, in the half-space $z>0$ with
boundary at $z=0$, provided the flux arising with the normal
component of $A^\mu$ satisfies the inequality
\begin{equation}
  {}^T u A^1 u \le 0.
  \label{eq:inequality}
\end{equation} 
This inequality determines the allowed boundary data for a well posed
initial-boundary value problem.

As expected from knowledge of  the characteristics of the wave equation, the
normal matrix $A^1$ has eigenvalues $0, \pm 1$, with the corresponding
eigenvectors 
\begin{eqnarray}
v_{0} = \left( \begin{array}{r} 1 \\ 0 \\ 0 \end{array} \right) , \quad
v_{\pm 1} = \left( \begin{array}{r} 0 \\ \mp 1 \\ 1\end{array} \right) .
\end{eqnarray}
These eigenvectors are associated with the variables 
\begin{equation}
h_0 = h , \quad h_{\pm 1} = f \mp k  =  h_{,z} \mp h_{,t} .
\end{equation}
Re-expressing the solution vector in terms of the eigenvectors,
\begin{equation}
   \tilde u = \left( \begin{array}{l} h_{-1} \\ h_0 \\ h_{+1} 
\end{array} \right)
= \left( \begin{array}{c} 
h_{,z} + h_{,t} 
\\ 
h 
\\ 
h_{,z} - h_{,t}  
\end{array} \right)
\end{equation}
the inequality (\ref{eq:inequality} implies that
homogeneous boundary data must take the form
\begin{equation}
          0 = h_{+1} - H h_{-1},
\end{equation}
where the parameter $H$ satisfies $H^2 \leq 1$.
The  component $h_{0}$, corresponding to the kernel of $A^1$, propagates
directly up the boundary and cannot be prescribed as boundary data. 

Non-homogeneous boundary data can be given in the form 
\begin{equation}
   q(t) = h_{+1} - H h_{-1} = h_{,z} (1-H) - h_{,t} (1+H) ,
\label{eq:bdrydata}
\end{equation}
where $q$ is an arbitrary function representing  the free boundary data at
$z=0$. Eq.~(\ref{eq:bdrydata}) shows how the scalar wave equation  accepts a
continuous range of boundary conditions. The well-known cases of Sommerfeld,
Dirichlet or Neumann boundary data are recovered by setting  $H = 0 , +1$, or,
$-1$, respectively. Note that there are consistency conditions at the edge
$(t=0,z=0)$. For instance, Dirichlet data corresponds to specifying $k=h_{,t}$
on the boundary and this must be consistent with the initial data for $k$.

\subsection{The reduced harmonic Einstein system}
\label{sec:hes}

The harmonic Einstein system consists of the six wave equations
(\ref{eq:gammawave}) and the four harmonic conditions (\ref{eq:auxeqs}). Since
the wave equations for $\gamma^{ij}$ are independent of the auxiliary variables
$\gamma^{t\alpha}$, the well-posedness of the initial-boundary value problem
for $\gamma^{ij}$ follows immediately. Furthermore, since the harmonic
conditions propagate the auxiliary variables up the boundary by ordinary
differential equations, the harmonic Einstein system has a well posed
initial-boundary value problem. A unique solution in the appropriate domain of
dependence is determined by the initial Cauchy data $\gamma^{ij}$ and
$\gamma^{ij}_{,t}$ at $t=0$, the initial data $\gamma^{t\alpha}$ at the edge
$t=z=0$ and the boundary data $\gamma^{ij}$ at $z=0$ given in any of  the forms
described in Appendix \ref{app:ricci} (e.g. Dirichlet, Neumann, Sommerfeld).

A solution of the linearized harmonic Einstein system satisfies
$\delta G^{ij}$ =0. As a result the Bianchi identities imply $\partial_t C^i =0$ and
$\partial_t C= -\partial_j C^j$ so that the constraints are satisfied provided
the constraint Eq's.~(\ref{eq:auxil1}) and (\ref{eq:auxil2}) are satisfied at
$t=0$.

The free boundary data for this system consists of six functions. However, as
shown in Ref. \cite{Friedrich98}, the vacuum Bianchi identities satisfied by
the Weyl tensor imply that only two independent pieces of Weyl data can be
freely specified at the boundary. We give the corresponding analysis for the
linearized Einstein system in Appendix \ref{app:weyl}. This makes it clear
that only two of the six pieces of metric boundary data are gauge invariant.
This is in accord with the four degrees of gauge freedom consisting of the
choice of linearized lapse $\delta \alpha =-h^{tt}/2$ (one free function) and
linearized shift $\delta \beta^i = -h^{ti}$ (three free functions).

A linearized evolution requires a unique lapse and shift, whose values can be
specified explicitly as space-time functions or specified implicitly in terms of
initial and boundary data subject to dynamical equations. In the case of a
harmonic gauge, in order to assess whether explicit space-time specification of
the lapse and shift is advantageous for the purpose of numerical evolution, it
is instructive to see how it affects the initial-boundary value problem for the
reduced harmonic Einstein system. In the linearized harmonic formulation
without boundary, a gauge transformation (\ref{eq:gauge}) to a shift-free gauge
$\gamma^{ti}=0$ is always possible within the domain of dependence of the
initial Cauchy data \cite{wald1984}. In the presence of a boundary, consider a
gauge transformation with $\xi^\alpha=(0,\xi^i)$, so that 
\begin{equation}
         \hat \gamma^{ti}= \gamma^{ti}+\partial^t \xi^i.
\end{equation}
For harmonic evolution of constrained initial data, both $\xi^i$ and
$\gamma^{ti}$ satisfy the wave equation. At $t=0$, choose Cauchy data for
$\xi^i$ satisfying
\begin{equation}
              0=\gamma^{ti}+\partial^t \xi^i 
\end{equation}
and
\begin{equation}
               0= \partial_t\gamma^{ti}+\partial_t \partial^t \xi^i 
	       =-\partial_j\gamma^{ji}-\nabla^2 \xi^i.
\end{equation}
On the boundary $z=0$, require that $\xi^i$ satisfy
\begin{equation}
              0=\gamma^{ti}+\partial^t \xi^i. 
\end{equation}
Then $\hat \gamma^{ti} =\gamma^{ti}+\partial^t \xi^i$ has vanishing Cauchy data
at $t=0$, vanishing Dirichlet boundary data at $z=0$ and satisfies the wave
equation, so that $\hat \gamma^{ti} =0$.  Thus, for evolution of constrained
data with a boundary,  a shift-free harmonic gauge is  possible. 

However, in this gauge, boundary data for all 6 components of $\gamma^{ij}$ can
no longer be freely specified since the harmonic condition implies 
\begin{equation}
        \partial_z \gamma^{zj}+\partial_A \gamma^{Aj} =0.
	\label{eq:noshift}
\end{equation}
This relates Neumann data for $\gamma^{zj}$ to Dirichlet data for  $\gamma^{AB}$
and would complicate any shift-free numerical evolution scheme. As an
example, one could freely specify Dirichlet boundary data for the 3 components
$\gamma^{AB}$ and (i) obtain Neumann boundary data $\partial_z \gamma^{zA}$ from
the $A$-components of Eq.~(\ref{eq:noshift}), (ii) evolve $\gamma^{zA}$ in terms
of initial Cauchy data and (iii) obtain Neumann boundary data $\partial_z
\gamma^{zz}$ from the $z$-component of Eq.~(\ref{eq:noshift}). Note the
nonlocality of step (ii), which would have to be carried out ``on the fly''
during a numerical evolution. 

The requirement that the shift vanish reduces the free boundary data to three
components. It is also possible to eliminate an additional free piece of data
by choosing a unit lapse, i.e. setting $h^{tt}=0$. Suppose the shift has been
set to zero, so that the harmonic condition implies $\partial_t
h^{tt}=-\partial_t h^i_i$. Consider a gauge transformation $\hat h^{tt}= 
h^{tt}-2\partial_t \xi^t$, where $\xi^\alpha=(\xi^t,\xi^i)$ satisfies
$\partial_t \xi^i=\partial^i \xi^t$ so that the shift remains zero.  For
harmonic evolution of constrained data, both $\xi^t$ and $h^{tt}$ satisfy the
wave equation. At $t=0$, choose Cauchy data for $\xi^t$ satisfying
\begin{equation}
              0=h^{tt}-2\partial_t \xi^t 
\end{equation}
and
\begin{equation}
             0=  \partial_t h^{tt}-2\partial^2_t  \xi^t 
	       = -\partial_t h^i_i -2\nabla^2 \xi^t
\end{equation}
(where we assume the Cauchy data is given on a non-compact set so that there
are no global obstructions to a solution $\xi^t$). On the boundary $z=0$,
require $\xi^t$ satisfy
\begin{equation}
            0=h^{tt}-2\partial_t \xi^t . 
\end{equation}
Then $\hat h^{tt}=0$ because it is a solution of the wave equation with
vanishing Cauchy data and Dirichlet boundary data. (Alternatively, the lapse
can be gauged to unity by a transformation satisfying $\Gamma^t =\Box \xi^t
=const$, so that the harmonic source function $\Gamma^t$ still drops out of 
Eq.~(\ref{eq:einstein}) for the Einstein tensor.)

A unit lapse and zero shift implies that $\partial_t\gamma^i_i =0$ so that
$\gamma^i_i$ cannot be freely specified at the boundary. Coupled with our
previous results, imposition of a unit lapse and zero-shift reduces the free
boundary data to the two trace-free transverse components
$\gamma^{AB}-(1/2)\delta^{AB}\gamma^C_C$, in accord with the two degrees of
gauge-free radiative freedom associated with the Weyl tensor. (In addition,
the initial Cauchy data must satisfy Eq.~(\ref{eq:noshift}),
$\partial_t\gamma^i_i =0$ and $\nabla^2 \gamma^i_i=0$). A similar result
arises in the study of the unit-lapse zero-shift initial-boundary value
problem for the linearized ADM equations \cite{cboun}. However, in the case of
harmonic evolution, it is clear that explicit specification of the lapse and
shift leads to a more complicated initial-boundary value problem. It is more
natural to retain the freedom of specifying 6 pieces of boundary data, which
then determine the lapse and shift implicitly during the course of the
evolution.

The reduction of the free boundary data can be accomplished by other gauge
conditions on the boundary, which are not directly based upon the lapse and
shift. An example, which plays a central role in the Friedrich-Nagy
formulation, is the specification of the mean curvature of the boundary. To
linearized accuracy, the unit outward normal to the boundary at $z=0$ is 
\begin{equation}
     N_\alpha \approx (1+\frac{1}{2}h_{zz})\nabla_\alpha z.
\end{equation}
The associated mean extrinsic curvature $(g^{\alpha\beta}-N^\alpha N^\beta
)\nabla_\beta  N_\alpha$ is given to linear order by
\begin{equation}
   \chi =  \partial_t h_{tz}-\partial_A h^A_z
	   +\frac{1}{2}(\partial_z h^A_A-\partial_z h_{tt}).
\end{equation}
Although the extrinsic curvature of a planar boundary vanishes in the
background Minkowski space, the linear perturbation of the background induces
a non-vanishing linearized extrinsic curvature tensor. 

Under a gauge transformation induced by the $\xi^\alpha$, the mean curvature
transforms according to
\begin{equation}
  \hat \chi =\chi+ 
    (\partial_t^2-\partial^A \partial_A)\xi^z.
\end{equation}
A gauge deformation of the boundary in the embedding space-time makes it
possible to obtain any mean curvature by solving a wave equation intrinsic to
the boundary. In this respect, the mean extrinsic curvature of the boundary is
pure ``boundary gauge'' and can be specified to eliminate one degree of gauge
freedom. When the harmonic conditions are satisfied, the mean curvature of the
boundary reduces to $\chi=\frac{1}{2}\partial_z h^{zz}$.

This discussion shows that there are various ways that the six free pieces of
boundary data can be restricted by gauge conditions. Such restrictions can be
important for an analytic understanding of the initial-boundary problem but
their usefulness for numerical simulation is a separate issue, especially in
applications where the boundary does not align with the numerical grid as
discussed in Sec. \ref{sec:implem}.

\subsection{The reduced harmonic Ricci system}
\label{sec:hrs}

The underlying equations of the reduced harmonic Ricci system are the six wave
equations (\ref{eq:hwave}) and the four harmonic conditions (\ref{eq:gauge_t})
and (\ref{eq:gauge_i}). The symmetric hyperbolic formulation of this system and
the analysis of its characteristics is given in Appendix \ref{app:ricci}. The
variables consist of $$\left(h^{ij}, T^{ij}, X^{ij}, Y^{ij}, Z^{ij}, h^{it},
\phi \right) ,$$ where $T^{ij}= \partial_t h^{ij}$,  $X^{ij}= \partial_x
h^{ij}$, $Y^{ij}= \partial_y h^{ij}$ and $Z^{ij}= \partial_z h^{ij}$. In a
well posed initial-boundary value problem, there are seven free functions that
may be specified at the boundary. For example, in the analogue of the Dirichlet
case, the free boundary data consists of $T^{ij}$ and $\phi$. 

In the context of the second order differential form given by 
Eq's.~(\ref{eq:hwave}), with the harmonic conditions (\ref{eq:gauge_t}) and
(\ref{eq:gauge_i}), the boundary data remains the same, e.g. Dirichlet data
$\partial_t h^{ij}$ and $\phi$. This determines the evolution of $h^{ij}$ by
the wave equation, which then provides source terms for the symmetric
hyperbolic subsystem (\ref{eq:gauge_t}) and (\ref{eq:gauge_i}). This subsystem
implies that $\Box \phi =0$, so that the evolution of $\phi$ is also governed
by the wave equation, with initial Cauchy data $\phi$ and $\partial_t \phi$ 
where $\partial_t \phi$ is provided by the initial Cauchy data $h^{it}$ via Eq.
(\ref{eq:gauge_i})). The evolution of $h^{it}$ is then obtained by integration
of Eq.~(\ref{eq:gauge_i}).

Unlike the reduced harmonic Einstein system where the constraints propagate up
the boundary by ordinary differential equations, the initial-boundary value
problem for the reduced harmonic Ricci system does not necessarily satisfy
Einstein's equations even if the constraints are initially satisfied. The
Bianchi identities (\ref{eq:momprop}) imply (\ref{eq:hamprop}) so that both $C$
and $\partial_t C$ would initially vanish but, since $C$ satisfies the wave
equation, $C$ would vanish throughout the evolution domain only if it vanished
on the boundary. In that case, Eq.~(\ref{eq:momprop}) would imply that the
momentum constraints were also satisfied throughout the evolution domain. 

Thus evolution of constrained initial data for the harmonic Ricci system
yields a solution of the Einstein equations if and only if the Hamiltonian
constraint is satisfied on the boundary. This is equivalent to requiring that
$\Box \phi =0$ on the boundary. If the evolution equations are satisfied,
we can express this in the form 
\begin{equation}
    \Box \phi = \frac {2}{3} \partial_t^2 H 
                   -\partial_i \partial_j H^{ij} =0,
\label{eq:boxphi}
\end{equation}
where $H=h^i_i$ and $H^{ij}=h^{ij}-\frac {1}{3}\delta^{ij} h^k_k$. This allows
formulation of the following well posed initial-boundary value problem for a
solution satisfying the constraints. 

We prescribe initial Cauchy data that satisfies the constraints for $H^{ij}$,
$H$, $\phi$ and $h^{it}$, and free boundary data for $H^{ij}$ and $\phi$. The
system of wave equations then determines $H^{ij}$. This allows integration of
Eq.~(\ref{eq:boxphi}) on the boundary to obtain Dirichlet boundary values for
determination of $H$ as a solution of the wave equation. The remaining fields
$\phi$ and $h^{it}$ are then determined as a symmetric hyperbolic subsystem.
Note that the boundary constraint (\ref{eq:boxphi}) reduces the free boundary
data from seven independent functions (for unconstrained solutions) to six, in
agreement with the free boundary data for solutions of the reduced harmonic
Einstein system. 

\section{Numerical implementation}
\label{sec:implem}

Numerical error is an essential new factor in the computational implementation
of the preceding analytic results. The initial Cauchy data cannot be expected
to obey the constraints exactly. In particular, machine roundoff error always
produces an essentially random component to the data which, in a linear
system,  evolves independently of the intended physical simulation. It is of
practical importance that a numerical evolution handle such random data without
producing exponential growth (and without an inordinate amount of numerical
damping). We designate as {\it robustly stable} an evolution code for the
linearized initial-boundary problem which does not exhibit exponential growth
for random (constraint violating) initial data and random boundary data.
This is the criterion previously used to establish robust stability for ADM evolution with
specified lapse and shift \cite{cboun}.

We test for robust stability using the 3-stage methodology proposed for
evolution-boundary codes in Ref. \cite{cboun}. The tests check that the
$\ell_\infty$ norm of the Hamiltonian constraint ${\cal C}$ does not exhibit
exponential growth under the following conditions.

\begin{itemize} 

\item {\it Stage I:} Evolution on a 3-torus with random initial
Cauchy data.

\item {\it Stage II:}  Evolution on a 2-torus with plane
boundaries, i.e.  $T^2\times[-L,L]$, with random initial Cauchy data and random
boundary data.

\item {\it Stage III:} Evolution with a cubic boundary with
random initial Cauchy data and random boundary data.

\end{itemize}

Stage I tests robust stability of the evolution code in the absence of a
boundary. Stage II tests robust stability of the boundary-evolution code for a
smooth boundary (topology $T^2$). Stage III tests robust stability of the
boundary-evolution code for a cubic boundary with faces, edges and corners, as
standard practice for computational grids based upon Cartesian coordinates.

We have established robust stability for the evolution-boundary codes,
described in Sec's. \ref{sec:hec} and \ref{sec:hrc}, the reduced harmonic
Einstein and Ricci systems. The tests are performed according to the procedures
outlined in Ref. \cite{cboun} for an evolution of 2000 crossing times (a time
of $4000L$, where $2L$ is the linear size of the computational domain) on a
uniform $48^3$ spatial grid with a time step $\Delta t =\Delta x/4$ (which
is slightly less than half the Courant-Friedrichs-Lewy limit and typical
of values used in numerical relativity).

For purposes such as singularity excision or Cauchy-characteristic
matching, there are computational strategies based upon spherical boundaries.
The extension of a robustly stable ADM boundary-evolution algorithm with given
lapse and shift from a cubic grid boundary to a spherical boundary is
problematic \cite{bela}. Robustly stable evolution-boundary algorithms for the
reduced harmonic Ricci and Einstein systems with spherical boundaries are
presented in Sec. \ref{sec:sphb}.

Numerical evolution of the fields $h^{ij}$ (or $\gamma^{ij}$) and $h^{tt}$ (or
$\gamma^{tt}$) is implemented on a uniform spatial grid  $(x_{[I]}, y_{[J]},
z_{[K]}) =  (I \, \Delta x, \, J \, \Delta x, \, K \, \Delta x)$, with time
levels $t^{[N]} = N \, \Delta t$. Thus a field component $f$ is represented by
its values $f^{[N]}_{[I,J,K]} = f(t^{[N]}, x_{[I]}, y_{[J]}, z_{[K]})$. In
order to obtain compact finite difference stencils for imposing  boundary
conditions, the fields $h^{it}$ (or $\gamma^{it}$) are represented on staggered
grids at staggered time-levels, where $f^{[N+1/2]}_{[I+1/2, J+1/2, K+1/2]}  =
f(t^{[N+1/2]}, x_{[I+1/2]}, y_{[J+1/2]},  z_{[K+1/2]})$.

The extensive tests of stability reported here were performed using a leapfrog
evolution algorithm in order not to bias the tests by introducing
excessive dissipation. We expect that the tests would also be satisfied
by more dissipative algorithms. This was borne out by a limited number of
evolutions using an iterative Crank-Nicholson algorithm (implemented as in
Ref. \cite{cboun}).  

\subsection{Robustly stable algorithms for the reduced harmonic Einstein system}
\label{sec:hec}

\centerline{{\bf Stage I}}
\medskip

Evolution of $\gamma^{ij}$ is done by a standard three-level leapfrog scheme,
with the wave equation in  second-differential-order form, i.e.
\begin{eqnarray}
\label{eq:fde.gamma^ij}
\frac{      \gamma^{ij[N+1]}_{[I,J,K]} 
     - 2 \, \gamma^{ij[N  ]}_{[I,J,K]}
          + \gamma^{ij[N-1]}_{[I,J,K]}
     }{(\Delta t)^2} &=& \gamma^{ij[N]}_{,xx[I,J,K]} 
                      +  \gamma^{ij[N]}_{,yy[I,J,K]} 
                      +  \gamma^{ij[N]}_{,zz[I,J,K]} \, .
\end{eqnarray}
Second spatial derivatives are computed via
\begin{equation}
\label{eq:fde.f_xx}
f_{,xx[I,J,K]} = 
\frac{ f_{[I+1,J,K]} - 2\, f_{[I,J,K]} + f_{[I-1,J,K]}}{(\Delta x)^2} \, .
\end{equation}
On a grid staggered in space and at staggered time levels, evolution of
$\gamma^{it}$ is carried out by the finite difference version of
Eq.~(\ref{eq:einstaux1}), i.e.
\begin{eqnarray}
\label{eq:fde.gamma^it}
\frac{ \gamma^{it[N+1/2]}_{[I+1/2,J+1/2,K+1/2]} 
     - \gamma^{it[N-1/2]}_{[I+1/2,J+1/2,K+1/2]}}{\Delta t} 
    &=&\\ 
     - \gamma^{ix[N]}_{,x[I+1/2,J+1/2,K+1/2]} 
     - \gamma^{iy[N]}_{,y[I+1/2,J+1/2,K+1/2]} 
    &-&\gamma^{iz[N]}_{,z[I+1/2,J+1/2,K+1/2]}.
                            \nonumber
\end{eqnarray}
Here first spatial derivatives are evaluated at the center of the integer grid
cells, i.e.
\begin{eqnarray}
\label{eq:fde.f_x}
f_{,x[I+1/2,J+1/2,K+1/2]} = 
 \frac{1}{4} \Big [
\frac{ f_{[I+1,J  ,K  ]} - f_{[I,J  ,K  ]}}{\Delta x} &+&
\frac{ f_{[I+1,J+1,K  ]} - f_{[I,J+1,K  ]}}{\Delta x} \\ +
\frac{ f_{[I+1,J  ,K+1]} - f_{[I,J  ,K+1]}}{\Delta x} &+&
\frac{ f_{[I+1,J+1,K+1]} - f_{[I,J+1,K+1]}}{\Delta x} \Big ]\, . \nonumber
\end{eqnarray}
Since $\gamma^{tt}$ is represented on the integer grid which is staggered in
space and time with respect to the grid representation of $\gamma^{it}$, the
finite difference equation used for updating $\gamma^{tt}$ is similar to
Eq.~(\ref{eq:fde.gamma^it}) used to update $\gamma^{it}$.

\centerline{{\bf Stage II}}
\medskip

The hierarchy of ten linearized equations for the reduced harmonic Einstein
system makes the initial-boundary value problem particularly simple to
implement as a finite difference algorithm. Let the plane boundary be given by
the $K_0$-th grid point,  with the Stage I evolution algorithm applied to all
points with $K < K_0$. Update of $\gamma^{ij}_{[K_0-1]}$ requires
$\gamma^{ij}_{[K_0]}$ as boundary data. The same boundary information allows
update of $\gamma^{it}_{[K_0 - 1/2]}$, which, in turn, allows update of
$\gamma^{tt}_{[K_0-1]}$ with no additional boundary data. However, it is
interesting to note that specification of free boundary data for
$\gamma^{tt}_{[K_0]}$  or $\gamma^{it}_{[K_0+1/2]}$ would not affect
stability simply because no evolution equation uses that data.

\centerline{{\bf Stage III}}
\medskip

Let the evolution domain be the cube $- L < x^i < L$ with the grid structured
so that the boundary lies on (non-staggered)  grid points, i.e., $x^i_{[1]} =
\pm L$, etc. The boundary data consist of $\gamma^{ij}$ on the faces, edges and
corners of the cube. The field $\gamma^{it}$ can then be updated at all
staggered grid points inside the cube, including those neighboring the boundary.
For instance, update of  $\gamma^{it}_{[3/2, 3/2,3/2]}$ 
involves use of  $\gamma^{ij}$ at the points 
${[\frac{3}{2} \pm \frac{1}{2},
\frac{3}{2} \pm \frac{1}{2},  \frac{3}{2} \pm \frac{1}{2}]}$, all of which are
on or inside the boundary. Similarly, evolution of $\gamma^{tt}$ can be carried
out at all interior grid points without further boundary data.

Robust stability of the evolution-boundary algorithm is demonstrated by the
graph of the Hamiltonian constraint in Fig.~\ref{fig:Stage.III}. The linear
growth results from momentum constraint violation in the initial data.

\subsection{Robustly stable algorithms for the reduced harmonic Ricci system}
\label{sec:hrc}

\centerline{{\bf Stage I}}
\medskip

Evolution of $h^{ij}$ is carried out identically as the  evolution of
 $\gamma^{ij}$ in the Einstein system (see Eq.~(\ref{eq:fde.gamma^ij})). The
fields $h^{ij}$ and $\phi$ are represented on the integer grid while $h^{it}$
is represented on a half-integer grid staggered in space and in time. Thus the
evolution equations (\ref{eq:gauge_t}) and (\ref{eq:gauge_i}) for $\phi$ and
$h^{it}$ have finite difference form 
\begin{eqnarray} \label{eq:fde.phi}
\frac{\phi^{[N+1]}_{[I,J,K]} -  \phi^{[N]}_{[I,J,K]}}{\Delta t}  + \left(
\partial_i h^{it} \right)^{[N+1/2]}_{[I,J,K]} + \frac{1}{2}
\delta_{ij}  \frac{h^{ij[N+1]}_{[I,J,K]} -  h^{ij[N]}_{[I,J,K]}}{\Delta t} &=&
0  \\ 
\label{eq:fde.h^it} \frac{
h^{it[N+1/2]}_{[I+1/2,J+1/2,K+1/2]}  -
h^{it[N-1/2]}_{[I+1/2,J+1/2,K+1/2]} }{\Delta t}
+ \left( \partial_i h^{it} - \frac{1}{2}  \delta_{jk} \partial^i h^{jk} +
\partial_j h^{ij} \right)^{[N]}_{[I+1/2,J+1/2,K+1/2]}
&=& 0 , \end{eqnarray} where the spatial derivative terms are computed
according to Eq.~(\ref{eq:fde.f_x}).

\centerline{{\bf Stage II}}
\medskip

{\it Unconstrained plane boundary}. As shown in Sec. \ref{sec:hrs}, seven free
functions can be prescribed as unconstrained boundary data for $\dot h^{ij}$
and $\phi$ in the reduced harmonic Ricci system. Again let the boundary be
defined by the $K_0$-th  grid point, with $K < K_0$ for interior points. Then
the boundary data $\dot h^{ij}_{[K_0]}$ and $\phi_{[K_0]}$ allows the evolution
algorithm to be applied to update $h^{ij}$ and $h^{it}$ at all interior points,
e.g. to update $h^{ij}_{[K_0-1]}$  and $h^{it}_{[K_0-1/2]}$, which in
turn allows update of $\phi$ at all interior points.

{\it Constrained plane boundary}. Conservation of the constraints in the
reduced harmonic Ricci system requires that the Hamiltonian constraint be
enforced at the boundary. In order to obtain a finite difference approximation
to a solution of Einstein's equations, the unconstrained evolution-boundary
algorithm must be modified to enforce the Hamiltonian constraint on the
boundary. On a $z=$const  boundary, we accomplish this, in accord with the
discussion in Sec. \ref{sec:hrs}, by prescribing freely the function $\phi$ and
the five components of  the traceless symmetric tensor $\partial_t H^{ij}$. The
missing ingredient,  $\partial_t H$ is updated at the boundary according to 
Eq.~(\ref{eq:boxphi}).

\begin{figure}
\centerline{\epsfxsize=4in\epsfbox{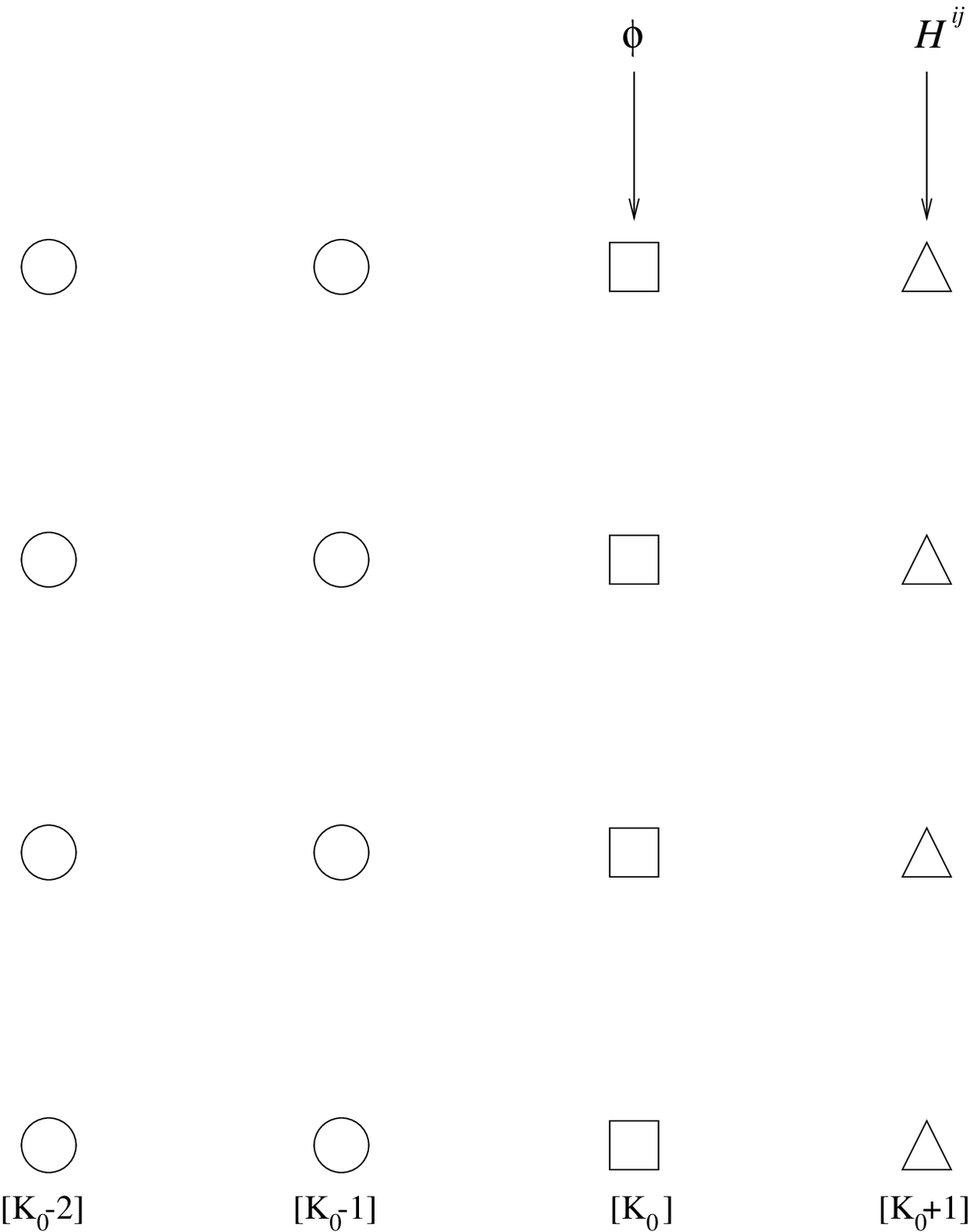}}
\caption{Boundary stencil for the Stage 2 
evolution-boundary algorithm of the Ricci system with constrained boundary.
Circles stand for interior grid points which are updated by the evolution
algorithm. The required boundary data for
$H^{ij}$ is provided  at the guard point $[K_0+1]$ (triangle), while
the boundary data for $\phi$ is provided at the boundary point $[K_0]$ 
(square). The Hamiltonian constraint is enforced at the 
boundary point $[K_0]$. }
\label{fig:stencil.II}
\end{figure}

In the finite difference algorithm, in order to be able to apply 
Eq.~(\ref{eq:boxphi}) via a centered three-point stencil, we introduce a
guard point at $z = z_{[K_0+1]}$, where the boundary data 
$\partial_t H^{ij}$ is provided. (See Fig.~\ref{fig:stencil.II}.)
Assuming that the fields $h^{ij}, h^{it}$ and $\phi$ are known
for $t \leq t^{[N]}$, $K \leq K_0$ and that
the boundary data $H^{ij}_{[K_0+1]}$ is known for $t \leq t^{[N]}$, 
we use the following evolution-boundary algorithm 
to compute these fields at $t^{[N+1]}$:
\begin{itemize}
\item (i) We update the fields $h^{ij}$ at time-level $t^{[N+1]}$ 
at all grid points within the numerical domain of dependence of the data known
at $t^{[N]}$, i.e., at all points which require no boundary data.
\item (ii) At the guard point $K_0+1$
we assign Dirichlet boundary data to the five independent components
of  the symmetric traceless $\left(\partial_t H^{ij}\right)^{[N+1/2]}_{[K_0+1]}$ 
by prescribing boundary
values for $\partial_t H^{xx}, \partial_t H^{xy}, \partial_t H^{xz}, 
\partial_t H^{yy}, \partial_t H^{yz}$ and setting  $\partial_t H^{zz} = -
\partial_t H^{xx} - \partial_t H^{yy}$. 
\item (iii) At the guard point $K_0+1$ we update the fields 
$H^{ij[N+1]}_{[K_0+1]} = H^{ij[N]}_{[K_0+1]} 
+ \left(\Delta t\right) \, \left(\partial_t H^{ij}\right)^{[N+1/2]}_{[K_0+1]}$.
\item (iv) At the boundary point $K_0$ we update 
$H^{ij[N+1]}_{[K_0]}$  using the field-equation 
$\Box H^{ij} = 0$, written in the finite-difference form
\begin{equation}
\label{eq:fde.box_Hij}
\frac{ H^{ij[N+1]}_{[K_0]} 
- 2 \, H^{ij[N]}_{[K_0]}
     + H^{ij[N-1]}_{[K_0]} }{\Delta t^2} = \left(\nabla^2 H^{ij}\right)^{[N]}_{[K_0]} .
\end{equation}
\item (v)  At the boundary point $K_0$
we compute the boundary values $H^{[N+1]}_{[K_0]}$
using the finite-difference version of Eq.~(\ref{eq:boxphi}), i.e.
\begin{equation}
\label{eq:fde.box_phi}
\frac{2}{3} \frac{ H^{[N+1]}_{[K_0]} 
            - 2 \, H^{[N]}_{[K_0]}
                 + H^{[N-1]}_{[K_0]} }{\Delta t^2}
                 -\left(\partial_i \partial_j H^{ij}\right)^{[N]}_{[K_0]}=0 .
\end{equation}
In Eq's.~(\ref{eq:fde.box_Hij}) - (\ref{eq:fde.box_phi})
all derivatives are computed using centered three-point expressions.
\item (vi) From knowledge of $H^{ij[N+1]}_{[K_0]}$ and $H^{[N+1]}_{[K_0]}$
we construct $h^{ij[N+1]}_{[K_0]}$.
\item (vii) We assign boundary data for $\phi^{[N+1]}_{[K_0]}$ and update  
 $\phi^{[N+1]}$  and  $h^{it[N+1/2]}$ according to 
Eq's.~(\ref{eq:fde.phi}) - (\ref{eq:fde.h^it}).
\end{itemize}

\centerline{{\bf Stage III}}
\medskip

{\it Unconstrained cubic boundary}. The unconstrained cubic boundary is
essentially the same as the unconstrained plane boundary, i.e., the functions
$\dot h^{ij}$ and $\phi$  are provided at the boundary. Then the evolution
equations are  applied to update $h^{ij}$, $h^{it}$ and $\phi$ at all interior
grid points. Recall that $h^{it}$ is represented on a staggered grid with some
interior points located a half grid-step away from the boundary. Nevertheless,
$h^{it}$ can be updated at such points using Eq.~(\ref{eq:fde.h^it}).

{\it Constrained cubic boundary}. The algorithm for enforcing the Hamiltonian 
constraint on a cubic boundary is an extension of the algorithm for a plane
boundary. The boundary functions $\partial_t H^{ij}$ are provided at a set of
guard points  that are $\Delta x$ outside the boundary of the cube, such that at
grid-points on the boundary of the cube we can use the field equations $\Box
H^{ij} = 0$ as well as  the boundary constraint Eq.~(\ref{eq:boxphi}) to update
$H^{ij}$ and $H$ at the boundary points. Boundary data for $\phi$ is provided at
grid points on the boundary of the cube, in a similar fashion to the constrained
plane boundary.

\begin{figure}
\centerline{\epsfxsize=4in\epsfbox{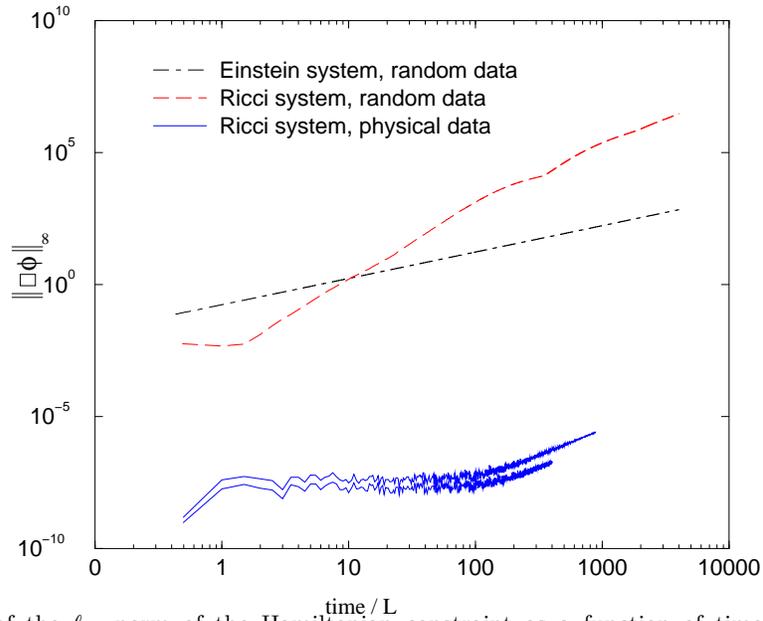}}
\caption{A log-log plot of the $\ell_\infty$ norm of the Hamiltonian constraint as
a function of time for a Stage 3 test of the evolution-boundary algorithm 
of the Einstein system and of the Ricci system with constrained boundary.
The upper two curves correspond to stability tests (random data),
while the lower two curves indicate performance tests (physical data).}
\label{fig:Stage.III}
\end{figure}

Robust stability of the evolution-boundary algorithm  for the constrained cubic
boundary is demonstrated by the graph of the Hamiltonian constraint in
Fig.~\ref{fig:Stage.III}.
In addition to the robust stability test results,
in Fig.~\ref{fig:Stage.III} we
have also included results from two physical runs,
based on the plane-wave solution
Eq's.~(\ref{eq:ctest.plwave-first}) - (\ref{eq:ctest.plwave-last}).
The longer physical run (up to $t/L \approx 1000 $) was performed with a
grid-size of $\Delta x = 1 / 80 $, while the shorter run (up to $t/L \approx
400 $) was performed with a grid-size of  $\Delta x = 1 / 120 $. Comparison of
the physical runs with different gridsizes indicates that the long-term
polynomial growth in the Hamiltonian constraint violation can be controlled by
grid-resolution.

\subsection{Spherical boundaries}
\label{sec:sphb}

The implementation of both the reduced harmonic Einstein and Ricci systems
display robust stability in Stage I, II and III tests. We now extend these
results to a a spherical boundary cut out of a Cartesian grid, with the
evolution domain defined by the interior of a sphere of radius $R$. In order
to update fields at these interior grid points, values are needed at a set of
``guard'' points consisting of grid points on or outside the boundary. Some of
these field values constitute free boundary data.   We extend our test of
robust stability to a fourth stage which checks that the $\ell_\infty$ norm of
the Hamiltonian constraint $C$ does not exhibit exponential growth under the
following condition.

\begin{itemize} 
\item {\it Stage IV:} Evolution with a spherical boundary with
random initial Cauchy data and random free boundary data at all
guard points.
\end{itemize}

As before, we perform the tests for an evolution of 2000
crossing times ($4000R$, where $2R$ is the diameter of the computational
domain) on a uniform $48^3$ spatial grid with a time step slightly less than
half the Courant-Friedrichs-Lewy limit. We have established Stage IV robust
stability for the following implementations of reduced harmonic Einstein and
Ricci revolution.

\subsubsection{The reduced harmonic Einstein system}

Let the evolution domain be defined by the interior of a sphere of radius $R$.
The update of $\gamma^{tt}$
at an interior point $[I,J,K]$ requires the values of
$\gamma^{it}$ at the points 
${[I \pm \frac{1}{2}, J \pm \frac{1}{2}, K \pm
\frac{1}{2}]}$, some of which  might be ``staggered-boundary'' points outside the
evolution domain. These staggered-boundary points are inside the spherical shell
 $R \leq \sqrt{x^2+y^2+z^2} < R + \frac{1}{2} \, \delta r_0$, 
if $\delta r_0 \geq \sqrt{3} \, \Delta x$.
At these points we update $\gamma^{it}$ by
the same finite-difference equation (\ref{eq:fde.gamma^it}) as for the
interior points. This defines a set of non-staggered boundary points for the field
$\gamma^{ij}$ determined by two conditions: (i) that the $\gamma^{ij}$ can be
updated at all interior points using Eq.~(\ref{eq:fde.gamma^ij}) and (ii)
that $\gamma^{it}$ can be updated at all interior and staggered-boundary points.
Both conditions are satisfied if we define the set of 
non-staggered boundary points 
by $R \leq \sqrt{x_{[I]}^2+y_{[J]}^2+z_{[K]}^2} < R + \delta r_0 $, 
where $\delta r_0 \geq \sqrt{3} \, \Delta x$.
At this set of  points we obtain values for 
$\gamma^{ij}$ via Eq.~(\ref{eq:fde.gamma^ij}), with 
  $\partial_t \gamma^{ij}$ provided in a ``guard-shell'' 
 $R + \delta r_0 \leq \sqrt{x^2+y^2+z^2} < R + \delta r_0 + \delta r_1$, 
where $\delta r_1 \geq \Delta x $.
The radius $R$ of the spherical boundary of the reduced harmonic Einstein system
is related to the linear size $2 L$ of the computational grid by
\begin{equation}
            R + \delta r_0 + \delta r_1 \leq L\,.
\end{equation}
This ensures that all guard points fall inside the domain $[-L, +L]^3$.

\subsubsection{The reduced harmonic Ricci system}

{\it Unconstrained spherical boundary}. 
The Stage IV unconstrained evolution-boundary algorithm for the reduced harmonic
Ricci system is similar to that of the reduced harmonic Einstein system.
We use two spherical shells, of thickness $\delta r_0$ and $\delta r_1$, with
$\delta r_0 \geq \sqrt{3}\, \Delta x$ and $\delta r_1 \geq \Delta x$.
The algorithm is the following:
\begin{itemize}
\item (i) We update all fields $h^{ij}$ at time-level $t^{[N+1]}$ at all 
non-staggered grid points
within the sphere of radius $R + \delta r_0$.
\item (ii) We provide boundary data $\left(\partial_t h\right)^{ij[N+1/2]}$ 
at all non-staggered
grid points within the spherical shell 
$R + \delta r_0 \leq \sqrt{x^2+y^2+z^2} < R + \delta r_0 + \delta r_1$. We update 
$h^{ij[N+1]} = h^{ij[N]} + \Delta t \, \left(\partial_t h\right)^{ij[N+1/2]}$ within 
the same spherical shell.
\item (iii) We provide boundary data $\phi^{[N+1]}$ within the spherical shell
$R \leq \sqrt{x^2+y^2+z^2} < R  + \delta r_0$.
\item (iv) We update  the field $\phi^{[N+1]}$ inside the sphere of radius $R$
according to Eq.~(\ref{eq:gauge_t}), and update the fields $h^{it[N+1/2]}$
inside the sphere of radius $R+\delta r_0/2$ according to Eq.~(\ref{eq:gauge_i}).
\end{itemize}

{\it Constrained spherical boundary}.  
The evolution-boundary algorithm for the  reduced harmonic
Ricci system with constrained spherical boundary is an extension of the algorithm
with unconstrained boundary.
In addition to the spherical shells defined 
in the case of the unconstrained boundary,
we define an additional set of guard points
$R + \delta r_0 + \delta r_1 \leq \sqrt{x_{[I]}^2+y_{[J]}^2+z_{[K]}^2} < 
R + \delta r_0 + \delta r_1+ \delta r_2$,
where the boundary data $\partial_t H^{ij}$ is provided.  The
quantity $\delta r_2$ is defined by two conditions: 
(i) the fields $H^{ij}$ can be updated
according to Eq.~(\ref{eq:fde.box_Hij})
at all non-staggered grid points 
$R + \delta r_0  \leq \sqrt{x_{[I]}^2+y_{[J]}^2+z_{[K]}^2} < 
R + \delta r_0 + \delta r_1$, and 
(ii) the Hamiltonian constraint Eq.~(\ref{eq:boxphi}) can be enforced 
at the same set of  grid points.
Both of these conditions are satisfied if $\delta r_2 \geq \sqrt{2} \, \Delta x$.
The evolution-boundary algorithm for the reduced harmonic Ricci system 
with a constrained spherical boundary is the following:
\begin{itemize}
\item (i) We update all fields $h^{ij}$ at time-level $t^{[N+1]}$ at 
all non-staggered grid points within the sphere or radius $R + \delta r_0$.
\item (ii) We update the fields $H^{ij[N+1]}$ 
at the set of boundary points 
$R + \delta r_0 \leq \sqrt{x_{[I]}^2+y_{[J]}^2+z_{[K]}^2} < 
R + \delta r_0 + \delta r_1$
via  Eq.~(\ref{eq:fde.box_Hij}). Then 
we update the field  $H^{[N+1]}$ at these points 
via  Eq.~(\ref{eq:fde.box_phi}). 
 From knowledge of $H^{ij[N+1]}$ and $H^{[N+1]}$
we construct $h^{ij[N+1]}$ at the same set of grid points.
\item (iii) We provide boundary data $\left(\partial_t H^{ij}\right)^{[N+1/2]}$ 
at the set of guard points
$R + \delta r_0 + \delta r_1 \leq \sqrt{x_{[I]}^2+y_{[J]}^2+z_{[K]}^2} < 
R + \delta r_0 + \delta r_1+ \delta r_2$, then we
update $H^{ij[N+1]} = H^{ij[N]} 
+ \Delta t \, \left(\partial_t H^{ij} \right)^{[N+1/2]}$  
at the same set of grid points.
\item (iv) We assign boundary data for $\phi^{[N+1]}$ at the set 
of boundary points
$R \leq \sqrt{x_{[I]}^2+y_{[J]}^2+z_{[K]}^2} < R + \delta r_0$.
\item (v) We update $\phi^{[N+1]}$ and $h^{it[N+1/2]}$ according to 
Eq's.~(\ref{eq:gauge_t}) - (\ref{eq:gauge_i}).
\end{itemize}

The radius $R$ of the constrained spherical
boundary for the reduced harmonic Ricci system and the linear size $2\, L$
of the computational grid are related by
\begin{equation}
R + \delta r_0 + \delta r_1 +  \delta r_2 \leq L.
\end{equation}
This ensures that all boundary points and all guard points fall inside the 
domain $[-L, +L]^3$.

\begin{figure}
\centerline{\epsfxsize=4in\epsfbox{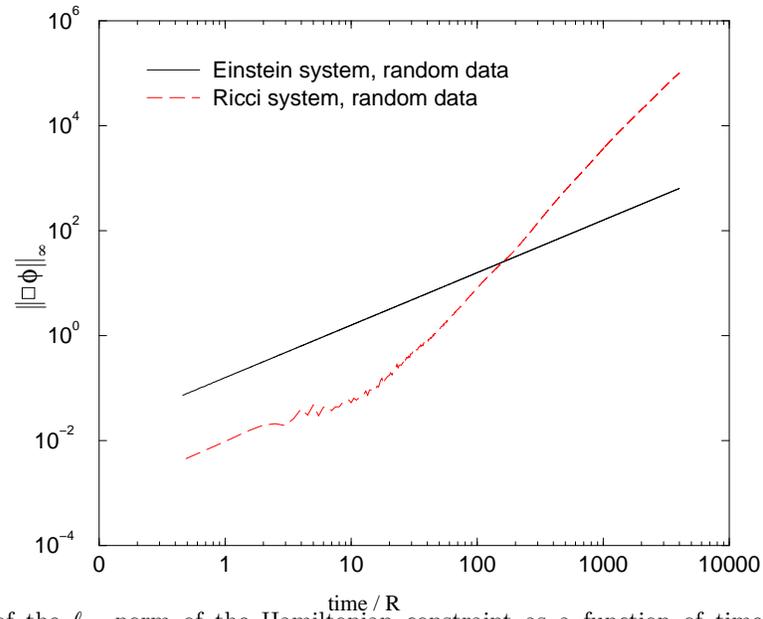}}
\caption{A log-log plot of the $\ell_\infty$ norm 
of the Hamiltonian constraint as
a function of time for a Stage 4 test of the evolution-boundary algorithm 
of the Einstein system and the Ricci system with constrained boundary.}
\label{fig:Stage.IV}
\end{figure}

The graphs of the Hamiltonian constraint in Fig.~\ref{fig:Stage.IV} illustrate
robust stability for a spherical  boundary for the reduced harmonic Einstein
system and for the reduced harmonic Ricci system with  constrained boundary
data. Comparison of Figs.~\ref{fig:Stage.III} - \ref{fig:Stage.IV} shows that
there is no significant difference between the Stage III and Stage IV
performances in terms of numerical stability.

\subsection{Convergence Tests}

In order to calibrate the performance of the algorithms we
carried out convergence tests based upon analytic solutions constructed from a
superpotential $\Phi^{\alpha \beta\mu \nu}$ symmetric in $(\alpha, \beta)$ and
$(\mu, \nu)$ and antisymmetric in $[\alpha, \mu]$, $[\beta, \nu]$, such that
$\Box \Phi^{\alpha \beta\mu \nu} = 0$. As a result of these symmetry
properties, the tensor  $\gamma^{\mu \nu} = \partial_\alpha \partial_\beta 
\Phi^{\alpha \beta\mu \nu}$ is symmetric and satisfies the linearized harmonic
Einstein equations $\Box \gamma^{\mu \nu} = 0$ and $\partial_\mu \gamma^{\mu
\nu} = 0$.

In our first testbed we choose  $\Phi^{\alpha \beta\mu \nu}$ as a
superposition of two solutions,
\begin{equation}
      \Phi^{\mu\mu \nu\nu} = A^{\mu \nu} + B^{\mu \nu},
      \quad \mu \neq \nu ,
\label{eq:ctest.plwave-first}
\end{equation}
with the remaining independent components of $\Phi^{\alpha \beta\mu \nu}$
set to zero. The solution $A^{\mu \nu}$ is defined by
\begin{eqnarray}
&& A^{ii} = A^{it} = A^{tt} = 0, \\
&& A^{ij} = \frac{8 \, a^{ij} \sin 
\left[ \omega_A \left( 
t -  \left(x^i+x^j\right) / \sqrt{2}
\right)\right]}{\omega_A^2}, \quad i \neq j
\end{eqnarray}
and $B^{\mu \nu}$ is defined by
\begin{eqnarray}
&&  B^{it} = \frac{8 \, b^{it} \sin \left[\omega_B 
\left(t - x^i\right)\right]  }
{\omega_B^2}, \quad B^{ij} = B^{tt} = 0.
\label{eq:ctest.plwave-last}
\end{eqnarray}

Here $A^{xy}$ is a  plane wave propagating  with frequency $\omega_A$ along
the diagonal of the$(x, y)$ plane, so that a wave crest leaving $x=-L$
travels a distance  $L \sqrt{2}$ before arriving at $x=+L$. Since the
topology of Stages I and  II imply periodicity in the $x$ direction, we set
$$
      \omega_A = \sqrt{2} \, \pi / L .
$$
Similarly, the frequency $\omega_B$ of the functions $B^{it}$
is set to
$$
\omega_B = \pi / L .
$$
In Stage III we use the same choices, while in the Stage IV tests we set
$$
      \omega_A = \sqrt{2} \, \pi / R , \quad \omega_B =  \pi / R .
$$
The amplitudes $a^{ij}, b^{it}$ were chosen to be
\begin{eqnarray}
\nonumber
a^{xy} = 1.1\cdot 10^{-8}, \quad
a^{xz} = 1.3\cdot 10^{-8}, \quad
a^{yz} = 1.2\cdot 10^{-8},
\\ \nonumber
b^{xt} = 1.4\cdot 10^{-8}, \quad
b^{yt} = 1.0\cdot 10^{-8}, \quad
b^{zt} = 1.5\cdot 10^{-8} .
\end{eqnarray}

Convergence runs used the plane wave solution. In Stages I - III we used the
grid sizes
$$
\frac{\Delta x}{2 \, L} = 
\frac{1}{80},  \; \frac{1}{100}, \; 
\frac{1}{120}, \; \frac{1}{160}  \, .
$$
while in Stage IV we used
$$
\frac{\Delta x}{2 \, R} = 
\frac{1}{80},  \; \frac{1}{100}, \; 
\frac{1}{120}, \; \frac{1}{160}  \, ,
$$
with the additional gridsize $\Delta x / (2 R) = 1/200$
in the case of the the reduced harmonic Ricci system with constrained spherical
boundary.

The time-step was set to $\Delta t = \Delta x / 4$.  
In the Stage IV test of the reduced harmonic Einstein system
the widths of the boundary shells were chosen to be $\delta r_0 = 1.8 \, \Delta x$
and $\delta r_1 = \Delta x$.  The same parameters were used when testing
the algorithm for the reduced harmonic Ricci system with unconstrained spherical
boundary.  The evolution-boundary algorithm for 
the reduced harmonic Ricci system with constrained spherical
boundary was tested using the parameters $\delta r_0 = 1.8 \, \Delta x, \;
\delta r_1 = 2.5 \Delta x$, and $\delta r_2 = 1.5 \, \Delta x$.

The code was used to
evolve the solutions from $t=0$ to $t=L$ ($t=R$ in Stage IV), at which time convergence was tested
by measuring  the $\ell_\infty$ and the $\ell_2$ 
norms of $\Box \gamma^{tt}$ for the Einstein system and of
$\Box \phi$ for the Ricci system, which test convergence
of the Hamiltonian constraint. The norms were evaluated
in the entire evolution domain.
In addition, we also checked convergence of the
metric components to their analytic values.

In addition to plane wave tests, we tested the qualitative performance
in Stage IV using an offset spherical wave based upon the
superpotential (with shifted origin)
\begin{eqnarray}
\Phi^{\mu \mu \nu \nu} = \frac{ f(t + \tilde r) - f(t - \tilde r)}
              {\tilde r},  \quad  \mu \neq \nu  ,
\end{eqnarray}
where $\tilde r = \sqrt{ x^2 + (y+a)^2 + z^2 }$ and
\begin{equation}
     f(u) = A \, \left(\frac{u}{w} \right) \exp \left[-\left(\frac{u}{w} 
            \right)^2\right] .
\end{equation}
The parameters $A, a$, and $w$ are set to 
$$
A = 2 \cdot w^3 \cdot 10^{-5}, \quad
a = 0.05 R, \quad w = 0.2 R .$$

\subsubsection{The reduced harmonic Einstein system}

Evolution requires Cauchy data at $t=0$ and boundary data at the guard points.
The Cauchy data   $\left\{\gamma^{ij}, \partial_t \gamma^{ij}, \gamma^{it},
\gamma^{tt} \right\}_{t=0}$ was provided by giving  $\gamma^{ij[0]},
\gamma^{ij[-1]}, \gamma^{tt[0]}$ and $\gamma^{it[1/2]}$ at all interior and
guard points. In addition, we provided  boundary data at each time-step by
giving  $\left( \partial_t \gamma^{ij} \right)^{[N+1/2]}$ at all guard points.
The metric and Hamiltonian constraint were found to be 2nd order convergent
for Stages I - IV.  In particular, in Stage IV, the norm of $\Box \gamma^{tt}$
vanished to  $O(\Delta^{1.99})$. 

\subsubsection{The reduced harmonic Ricci system}

In the case of the Ricci system the 
Cauchy data   $\left\{h^{ij}, \partial_t h^{ij}, h^{it}, \phi
\right\}_{t=0}$ was provided by giving $h^{ij[0]}, h^{ij[-1]}, \phi^{[0]}$ and
$h^{it[1/2]}$ at all interior and boundary points.  In addition,
when the Hamiltonian constraint was numerically imposed at the boundary, 
we also provided $\left\{H^{ij}\right\}_{t=0}$ by giving $H^{ij[0]}$.

We first tested the code without numerically imposing  the Hamiltonian
constraint. In this case we provided  boundary data at each time-step by
giving  $\left( \partial_t h^{ij} \right)^{[N+1/2]}$ and $\phi^{[N]}$
at all guard points.

Next we tested the code with the Hamiltonian constraint numerically imposed
at the boundary. Thus we first prescribed the traceless  $\left( \partial_t
H^{ij} \right)^{[N+1/2]}$ and $\phi^{[N]}$ at all guard points, then
computed $H$ at each time-step via the boundary constraint
Eq.~(\ref{eq:boxphi}). 

In all cases we found the numerically evolved
metric functions converge to their analytic values to $O(\Delta^2)$.
In Stage I, $\Box \phi$ vanished to roundoff accuracy,
while in Stages II - III it vanished to second order accuracy. In particular, 
for the Stage III algorithm with constrained boundary, 
$\Box \phi $ converged to zero as $O(\Delta^{1.99})$.

In Stage IV, with constrained boundary, we found that the $\ell_2$ norm of $\Box
\phi$ vanishes to first order accuracy. However, the $\ell_\infty$  norm
decreases linearly with grid size only for $\Delta x / (2 R) = 1 / 80, 1/100$ and
$ 1/120$ but fails to show further decrease for $\Delta x / (2 R) = 1 / 160$ and
$ 1/200$. This anomalous behavior of the $\ell_\infty$ norm stems from the random
way in which guard points are required at different sites near the boundary. This
introduces an unavoidable nonsmoothness to the second order error in the metric
components, which in turn leads to $O(1)$ error in the second spatial derivatives
occurring in $\Box \phi$ or in the Hamiltonian constraint. Unlike the Einstein
system in which the constraints propagate tangent to the boundary, this error in
the Ricci system propagates along the light cone into the interior. However,
since its origin is a thin boundary shell whose width is $O(\Delta x)$, the
$\ell_2$ norm of $\Box \phi$ remains convergent to first order. We expect that
the convergence of the Hamiltonian constraint for a spherical boundary would be
improved by matching the interior solution on the Cartesian grid to an exterior
solution on a spherical grid aligned with the boundary, as is standard practice
in treating irregular shaped boundaries.

\subsubsection{Simulation of an outgoing wave using a constrained spherical
boundary}

\begin{figure}
\centerline{\epsfxsize=2.5in\epsfbox{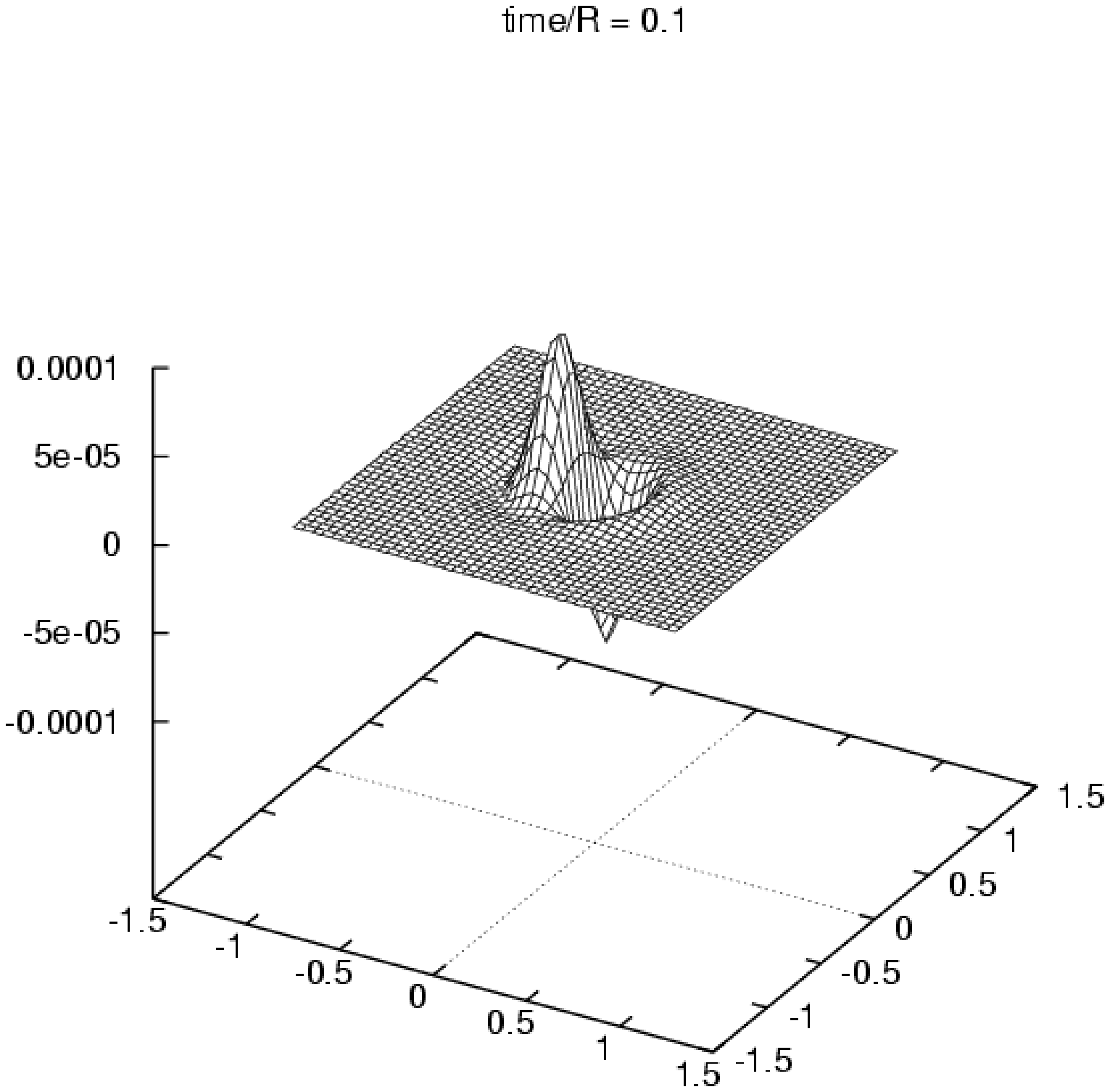}}
\vspace{.2in}
\centerline{\epsfxsize=2.5in\epsfbox{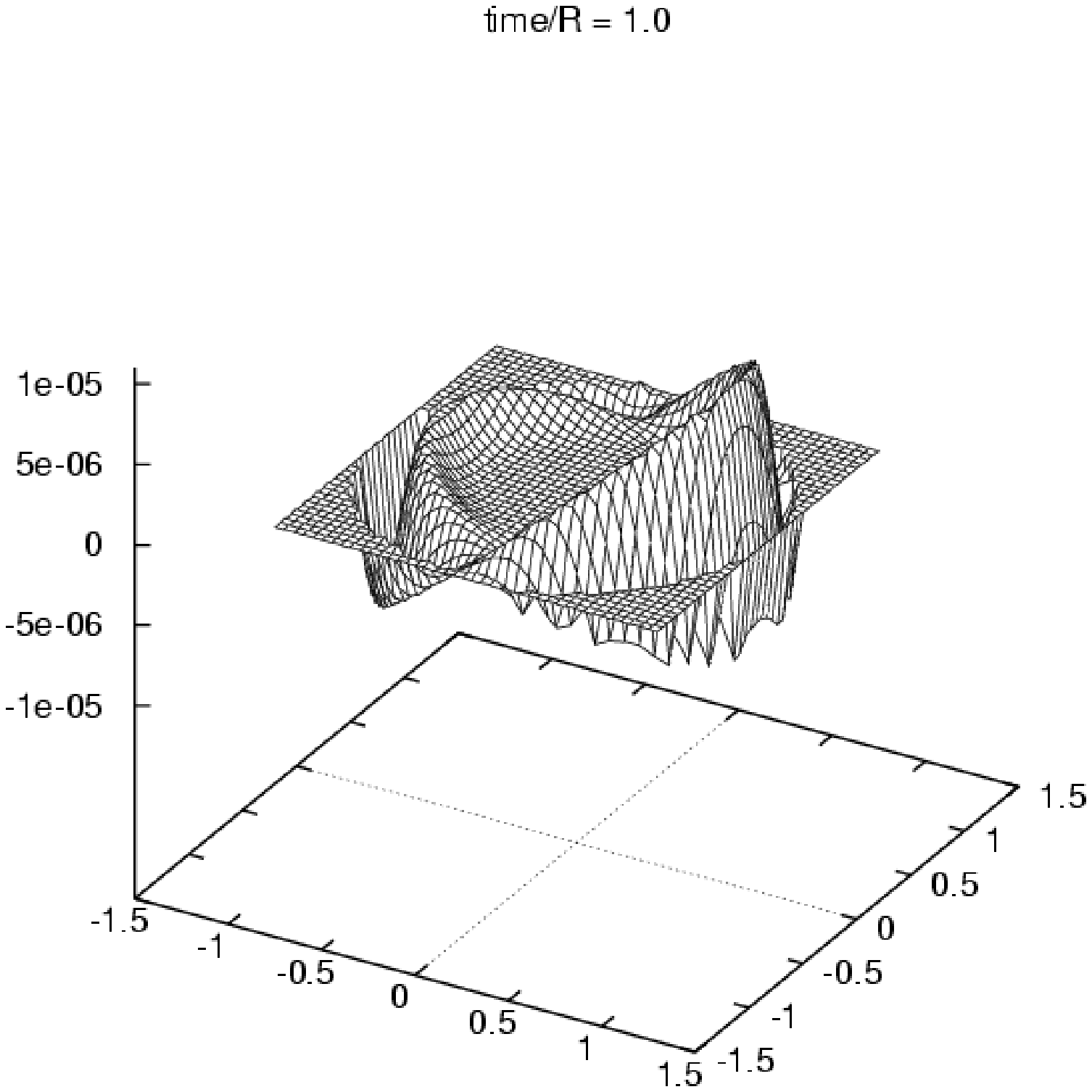}}
\vspace{.2in}
\centerline{\epsfxsize=2.5in\epsfbox{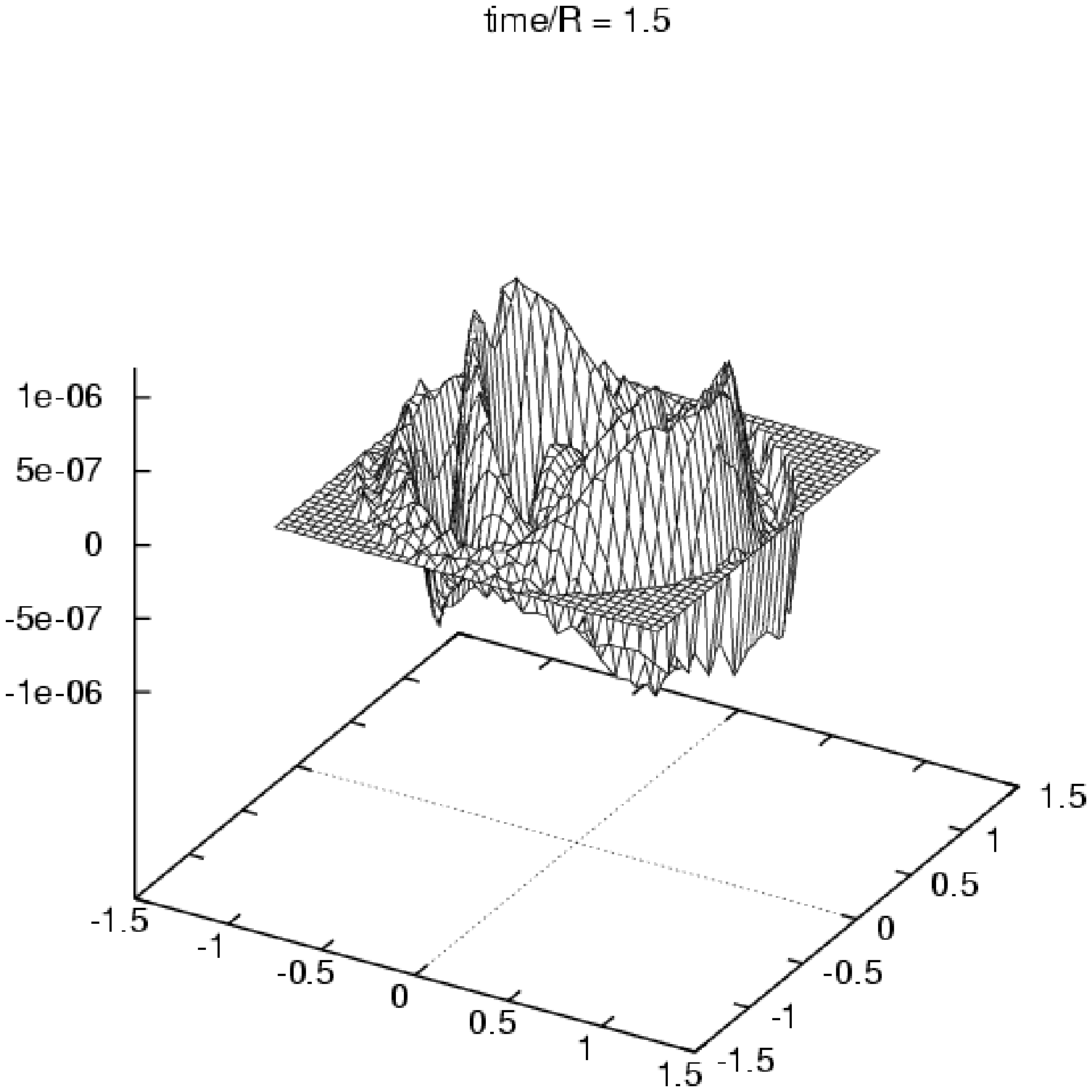}}
\caption{Stage IV evolution of an outgoing solution using the 
reduced harmonic Ricci system with a constrained spherical boundary.
The plots show the metric field $h^{yt}$ at $x=0$, for $t/R=0.1$ (top),
$t/R = 1.0$ (middle) and $t/R = 1.5$ (bottom). In the bottom plot, the field
has decayed by two orders of magnitude.}
\label{fig:2Dplots}
\end{figure}

We also tested the code's ability to evolve an outgoing spherical wave
traveling off center with respect to a spherical boundary of radius $R$.
Fig.~\ref{fig:2Dplots} illustrates a simulation performed using the Stage IV
algorithm for the reduced harmonic Ricci system, with the Hamiltonian
constraint numerically enforced at the boundary. The  metric fields were
evolved from $t=0$ to $t=1.5 R$, using a grid of $\Delta x / (2 R) = 1/120$.
After the analytic wave has propagated out of the computational domain, the
remnant error is two orders of magnitude smaller than the initial signal. This
shows that artificial reflection off the boundary is well controlled even in
the computationally challenging case of a piecewise cubic spherical boundary.

\acknowledgements

We thank Helmut Friedrich for numerous discussions of the initial-boundary
value problem. Our numerical code was based upon an earlier collaboration with
Roberto G\'{o}mez. This work has been partially supported by NSF grant PHY
9988663 to the University of Pittsburgh. Computer time was provided by the
Pittsburgh Supercomputing Center and by NPACI.

\appendix

\section{Hyperbolic formulation of the linearized harmonic Ricci system}
\label{app:ricci}

In order to study the evolution of the system
consisting of Eq's.~(\ref{eq:hwave}), (\ref{eq:gauge_t}), 
and (\ref{eq:gauge_i}) in the half-space $z>0$ with
boundary at $z=0$,
we employ the auxiliary variables $
T^{ij} = \partial_t h^{ij},
X^{ij} = \partial_x h^{ij},
Y^{ij} = \partial_y h^{ij},
Z^{ij} = \partial_z h^{ij},
\phi=\frac{1}{2}h^{tt}$. 
In terms of the variables
$$\left(h^{ij}, T^{ij}, X^{ij}, Y^{ij}, Z^{ij}, h^{it}, \phi \right) ,$$
the system takes the form
\begin{eqnarray}
\label{eq:ev_system_first}
\partial_t h^{ij} &=& T^{ij}\label{eq:h_eq}\\
\partial_t T^{ij} &=& \partial_x X^{ij} + \partial_y Y^{ij} + \partial_z Z^{ij}
\label{eq:T_eq}\\
\partial_t X^{ij} &=& \partial_x T^{ij}\label{eq:X_eq}\\
\partial_t Y^{ij} &=& \partial_y T^{ij}\label{eq:Y_eq}\\
\partial_t Z^{ij} &=& \partial_z T^{ij}\label{eq:Z_eq}\\
\partial_t h^{x t} &=& - \partial_x \phi
- \frac{1}{2} \, X^{xx} 
+ \frac{1}{2} \, X^{yy} 
+ \frac{1}{2} \, X^{zz} - Y^{x y} - Z^{x z}\label{eq:h_xt_eq}\\
\partial_t h^{y t} &=& - \partial_y \phi
+ \frac{1}{2} \, Y^{xx} 
- \frac{1}{2} \, Y^{yy} 
+ \frac{1}{2} \, Y^{zz} 
- X^{xy} - Z^{yz}\label{eq:h_yt_eq}\\
\partial_t h^{z t} &=& - \partial_z \phi
+ \frac{1}{2} \, Z^{xx} 
+ \frac{1}{2} \, Z^{yy} 
- \frac{1}{2} \, Z^{zz} 
- X^{xz} - Y^{yz} \label{eq:h_zt_eq}\\
\partial_t \phi &=& - \partial_x h^{xt} - \partial_y h^{yt} - \partial_z h^{zt}
- \frac{1}{2} \, T^{xx}- \frac{1}{2} \, T^{yy}- \frac{1}{2} \, T^{zz} .
\label{eq:phi_eq}
\label{eq:ev_system_last}
\end{eqnarray}
Next we define the 34-dimensional vector $u$ by 
\begin{equation}
u = {}^{T}(
      h^{xx}, h^{xy}, h^{xz}, h^{yy}, h^{yz}, h^{zz},  
      T^{xx}, \ldots, T^{zz}, 
      X^{xx}, \ldots, X^{zz}, 
      Y^{xx}, \ldots, Y^{zz}, 
      Z^{xx}, \ldots, Z^{zz},
      h^{xt}, h^{yt}, h^{zt}, \phi ).
\label{eq:udef1}
\end{equation}
The system of equations 
(\ref{eq:ev_system_first}) - (\ref{eq:ev_system_last}) then has the form 
\begin{equation}
A^\mu \, \partial_\mu u = B\,u 
\end{equation}
where $A^t = {\bf I}_{34\times34}$ is the identity matrix,
\begin{equation}
A^z = \left( 
\begin{array}{ccccccccc}
0_{6\times 6} 
  & 0_{6\times 6} 
      & 0_{6\times 6} 
          & 0_{6\times 6} 
              & 0_{6\times 6} 
                  & 0_{1 \times 6} 
                      & 0_{1\times 6} 
                          & 0_{1\times 6}
                              & 0_{1\times 6} \\
0_{6\times 6} 
  & 0_{6\times 6} 
      & 0_{6\times 6} 
          & 0_{6\times 6} 
              & -{\bf I}_{6 \times 6} 
                  & 0_{1 \times 6} 
                      & 0_{1\times 6} 
                          & 0_{1\times 6} 
                              & 0_{1\times 6} \\
0_{6\times 6} 
  & 0_{6\times 6} 
      & 0_{6\times 6} 
          & 0_{6\times 6} 
              & 0_{6\times 6} 
                  & 0_{1 \times 6} 
                      & 0_{1\times 6} 
                          & 0_{1\times 6}
                              & 0_{1\times 6} \\
0_{6\times 6} 
  & 0_{6\times 6} 
      & 0_{6\times 6} 
          & 0_{6\times 6} 
              & 0_{6\times 6} 
                  & 0_{1 \times 6} 
                      & 0_{1\times 6} 
                          & 0_{1\times 6}
                              & 0_{1\times 6} \\
0_{6 \times 6} 
  & -{\bf I}_{6 \times 6} 
      & 0_{6 \times 6} 
          & 0_{6 \times 6} 
              & 0_{6 \times 6} 
                  & 0_{1 \times 6} 
                      & 0_{1 \times 6} 
                          & 0_{1 \times 6} 
                              & 0_{1 \times 6} \\
0_{6\times 1} 
  & 0_{6\times 1} 
      & 0_{6\times 1} 
          & 0_{6\times 1} 
              & 0_{6\times 1} 
                  & 0_{1\times 1} 
                      & 0_{1\times 1} 
                          & 0_{1\times 1} 
                              & 0_{1\times 1} \\
0_{6\times 1} 
  & 0_{6\times 1} 
      & 0_{6\times 1} 
          & 0_{6\times 1} 
              & 0_{6\times 1} 
                  & 0_{1 \times 1} 
                      & 0_{1\times 1} 
                          & 0_{1\times 1}
                              & 0_{1\times 1} \\
0_{6\times 1} 
  & 0_{6\times 1} 
      & 0_{6\times 1} 
          & 0_{6\times 1} 
              & 0_{6\times 1} 
                  & 0_{1 \times 1} 
                      & 0_{1\times 1} 
                          & 0_{1\times 1}
                              & + {\bf I}_{1\times 1} \\
0_{6\times 1} 
  & 0_{6\times 1} 
      & 0_{6\times 1} 
          & 0_{6\times 1} 
              & 0_{6\times 1} 
                  & 0_{1\times 1} 
                      & 0_{1\times 1} 
                          & + {\bf I}_{1\times 1} & 0_{1\times 1}\\
\end{array}
\right)
\end{equation}
and 
\begin{equation}
B = \left(
\begin{array}{ccccccccc}
0_{6\times 6} 
  & {\bf I}_{6 \times 6} 
      & 0_{6\times 6} 
           & 0_{6\times 6} 
               & 0_{6 \times 6} 
                  & 0_{1 \times 6} 
                      & 0_{1 \times 6} 
                          & 0_{1 \times 6} 
                              & 0_{1\times 6} \\
0_{6 \times 6} 
  & 0_{6 \times 6} 
      & 0_{6 \times 6} 
          & 0_{6 \times 6} 
              & 0_{6 \times 6} 
                  & 0_{1 \times 6} 
                      & 0_{1 \times 6} 
                          & 0_{1 \times 6} 
                              & 0_{1 \times 6} \\
0_{6 \times 6} 
  & 0_{6 \times 6} 
      & 0_{6 \times 6} 
          & 0_{6 \times 6} 
              & 0_{6 \times 6} 
                  & 0_{1 \times 6} 
                      & 0_{1 \times 6} 
                          & 0_{1 \times 6} 
                              & 0_{1 \times 6} \\
0_{6 \times 6} 
  & 0_{6 \times 6} 
      & 0_{6 \times 6} 
          & 0_{6 \times 6} 
              & 0_{6 \times 6} 
                  & 0_{1 \times 6} 
                      & 0_{1 \times 6} 
                          & 0_{1 \times 6} 
                              & 0_{1 \times 6} \\
0_{6 \times 6} 
  & 0_{6 \times 6} 
      & 0_{6 \times 6} 
          & 0_{6 \times 6} 
              & 0_{6 \times 6} 
                  & 0_{1 \times 6} 
                      & 0_{1 \times 6} 
                          & 0_{1 \times 6} 
                              & 0_{1 \times 6} \\
0_{6\times 1} 
  & 0_{6\times 1}
      & {\bf C}^x
          & - \delta^{2j} 
              & - \delta^{3j} 
                  & 0_{1\times 1} 
                      & 0_{1\times 1} 
                          & 0_{1\times 1} 
                              & 0_{1\times 1} \\
0_{6\times 1} 
  & 0_{6\times 1} 
      & - \delta^{2j} 
          & {\bf C}^y 
              & - \delta^{5j} 
                  & 0_{1\times 1} 
                      & 0_{1\times 1} 
                          & 0_{1\times 1} 
                              & 0_{1\times 1} \\
0_{6\times 1} 
  & 0_{6\times 1} 
      & - \delta^{3j} 
          & - \delta^{5j} 
              & {\bf C}^z 
                  & 0_{1\times 1} 
                      & 0_{1\times 1} 
                          & 0_{1\times 1} 
                              & 0_{1\times 1} \\
0_{6\times 1} 
  & {\bf C}^t 
      & 0_{6\times 1} 
          & 0_{6\times 1} 
              & 0_{6\times 1} 
                  & 0_{1\times 1} 
                      & 0_{1\times 1} 
                          & 0_{1\times 1} 
                              & 0_{1\times 1} \\
\end{array}
\right)
\end{equation}
where
\begin{eqnarray}
{\bf C}^x &=& \frac{1}{2}\left(-1, 0, 0, +1, 0, +1 \right), \\
{\bf C}^y &=& \frac{1}{2}\left(+1, 0, 0, -1, 0, +1 \right), \\
{\bf C}^z &=& \frac{1}{2}\left(+1, 0, 0, +1, 0, -1 \right), \\
{\bf C}^t &=& \frac{1}{2}\left(-1, 0, 0, -1, 0, -1 \right) .
\end{eqnarray}

The matrix $A^z$ has the eigenvalue $+1$, with multiplicity
$7$ and eigenvectors 
$$
\left( \partial_t - \partial_z \right) h^{ij}, \quad \phi + h^{zt} ; 
$$
the eigenvalue $-1$, with multiplicity $7$ and eigenvectors
$$
\left( \partial_t + \partial_z \right) h^{ij}, \quad \phi - h^{zt} ;
$$
and the kernel of the matrix has dimension $20$, with 
a basis
$$
h^{ij}, \quad \partial_x h^{ij}, \quad \partial_y h^{ij}, \quad
h^{xt}, h^{yt} .
$$
In the eigen-basis defined by $A^z$ the vector $u$ defined in 
Eq.~(\ref{eq:udef1}) takes the form
\begin{equation}
u = {}^T(u_{-}, u_{0}, u_{+} )
\end{equation}
with
\begin{eqnarray}
u_{-} &=& {}^T( 
     T^{xx} + Z^{xx}, \ldots, T^{zz} + Z^{zz}, \phi - h^{zt} ) 
\\
u_{0} &=& {}^T( 
     h^{xx}, \ldots, h^{zz}, 
     X^{xx}, \ldots, X^{zz}, 
     Y^{xx}, \ldots, Y^{zz}, h^{xt}, h^{yt} )
\\
u_{+} &=& {}^T( 
     T^{xx} - Z^{xx}, \ldots, T^{zz} - Z^{zz}, \phi + h^{zt} ) .
\end{eqnarray}
Non-homogeneous boundary data can be given in terms of a free column vector
field $q$
in the form
\begin{equation}
    q = u_{+} - H u_{-}
\end{equation}
where $H$ can be any $7 \times 7$ matrix satisfying
\begin{equation}
 - \, {}^T a \, a + {}^T a \, {}^T H \, H \, a \leq 0, \quad
a \in {\bf R}^7 .
\label{eq:Hcond}
\end{equation}
The three simplest matrices that satisfy the condition (\ref{eq:Hcond})
are $- {\bf I}_{7 \times 7}, 0_{7 \times 7}$ and ${\bf I}_{7 \times 7}$ .
The first of these corresponds to specifying Neumann data for $h^{ij}$
and Dirichlet data for $h^{zt}$. Using the zero matrix as a candidate for
$H$ corresponds to giving Sommerfeld data for $h^{ij}$ and specifying 
the quantity $\phi - h^{zt}$. Last, picking $H$ to be the identity 
matrix corresponds to giving  Dirichlet data for $\partial_t h^{ij}$ as well
as for $\phi$. Note that the evolution system 
(\ref{eq:ev_system_first}) - (\ref{eq:ev_system_last})
accepts a much richer class of boundary conditions than the three
we just mentioned. One simply needs to pick a matrix $H$ that satisfies
Eq.~(\ref{eq:Hcond}) and the choice of $H$ defines the seven free functions
that are to be specified at the boundary.

\section{Weyl data on boundary}
\label{app:weyl}

The curvature tensor, which provides gauge invariant fields, decomposes
into the Ricci curvature, which vanishes if the evolution and constraint
equations are satisfied, and the Weyl curvature $C_{\alpha\beta\gamma\delta}$.
In order to analyze the
boundary freedom, it is convenient make the following choice of a
complete, independent set of 10 linearized Weyl tensor components
$K_{\alpha\beta\gamma\delta} =\delta C_{\alpha\beta\gamma\delta}$:
$C_{AB}=K_{tABt}-\frac{1}{2}\delta_{AB}\delta^{CD}K_{tCDt}$,
$K_t{}^A{}_{At}$, $K_{tAzt}$, $K_{tzxy}$, $K_t{}^A{}_{AB}$,
and $D_{AB}=K_{tABz}-\frac{1}{2}\delta_{AB}\delta^{CD}_{tCDz}$.
We use the linearized vacuum Bianchi identities $\partial_{[\alpha}
K_{\beta\gamma]\mu\nu}=0$ and $\partial^\delta K_{\alpha\beta\gamma\delta}=0$
to show that the  Weyl data which can be freely specified on the boundary can
be reduced to the 2 independent components $C_{AB}$. 

First, the identity $\partial^\delta K_{tzz\delta}=0$ implies
(after using the trace-free property of the Weyl tensor)
\begin{equation}
      \partial_t K_t {}^A{}_{At} -\partial^B K_t{}^{A}{}_{AB} =0
      \label{eq:trac}
\end{equation}
which determines the boundary behavior of $C_t{}^A{}_{At}$
in terms of the remaining 9 Weyl components.

Next, note that the identity 
\begin{equation}
   \partial_t K_{tABC}+\partial_B K_{tACt}+\partial_C K_{tAtB} =0.
    \label{eq:ctprop}
\end{equation}
implies 
\begin{equation}
      \partial_t K_t{}^A{}_{AC} =-\partial^A C_{AC}+ 
        +\frac{1}{2} \partial_C K_t{}^D{}_{Dt},
\end{equation}
or, taking a $t$-derivative and using Eq.~(\ref{eq:trac}), that
\begin{equation}
      \partial_t^2 K_t{}^A{}_{AC} =-\partial_t\partial^A C_{AC} 
        +\frac{1}{2} \partial_C \partial^D K_t{}^A{}_{AD}.
\end{equation}
This gives a propagation equation intrinsic to the boundary which determines
the time dependence of $K_t{}^A{}_{AC}$ in terms of the boundary data for
$C_{AB}$. (Note that $K_t{}^A{}_{AC}$ propagates up the boundary
with velocity $0$ in one mode and in a cone with velocity $1/\sqrt{2}$
in the other mode.)

Next, the identity 
\begin{equation}
   \partial^\delta K_{tAz\delta}
        =-\partial^t K_{tAzt} +\partial^B K_{tAzB} =0
\end{equation}
determines the time dependence of $C_{tAzt}$; and the identity
\begin{equation}
   \partial_t K_{tzAB}+\partial_B K_{tztA}+\partial_A K_{tzBt} =0
\end{equation}
determines the time dependence of $K_{tzxy}$. This reduces the free Weyl data
on the boundary to the 4 independent components $C_{AB}$ and $D_{AB}$.

However, the specification of $D_{AB}$, in addition to $C_{AB}$, would lead
to an inconsistent boundary value problem. This can be seen from
the identity 
\begin{equation}
 \partial^\delta K_{tAB\delta}
    =-\partial_t K_{tABt}+\partial_z K_{tABz}+\partial^C K_{tABC}=0,
\end{equation}
which determines Neumann data for $D_{AB}$ in terms of Dirichlet
data for $C_{AB}$ and other known quantities. Similarly, the identity
\begin{equation}
   \partial_t K_{tABz}-\partial_z K_{tABt}+\partial_B K_{tAzt} =0
   \label{eq:bianch}
\end{equation}
determines Neumann data for $C_{AB}$. Thus, since the components of the
Weyl tensor satisfy the wave equation, the specification of both $C_{AB}$ and
$D_{AB}$ as free Dirichlet boundary data leads to an inconsistent boundary
value problem.

The determination of boundary boundary values for $D_{AB}$ from boundary data
for $C_{AB}$ is a global problem which first requires solving the wave equation
to determine $C_{AB}$ from its boundary and initial data. Then the time
derivative of the trace-free part of Eq.~(\ref{eq:bianch}) yields
\begin{equation}
   \partial_t^2 D_{AB}-\partial_B \partial^C D_{AC}
              +\frac{1}{2} \delta_{AB}\partial^C \partial^D D_{CD}
    -\partial_t \partial_z C_{AB} =0
\end{equation}
which propagates $D_{AB}$ up the boundary in terms of initial data.
Defining $C=q^A q^B C_{AB}$ and  $D=q^A q^B D_{AB}$, with
$q^A\partial_A=\partial_x +i\partial_y$, this reduces to
\begin{equation}
   \partial_t^2 D -\frac{1}{2} \partial_A \partial^A D
    -\partial_t \partial_z C =0,
\end{equation}
which has propagation velocity $1/\sqrt{2}$. (Note that this is but one of the
variations consistent with the maximally dissipative condition used by
Friedrich and Nagy \cite{Friedrich98}. In the case of unit lapse and vanishing
shift, assigning boundary data for $C$ is equivalent to assigning data for the
trace-free part of the intrinsic 2-metric of the boundary foliation,
consistent with results found in Ref. \cite{cboun}.)

\newpage

\newpage

\newpage

\end{document}